\documentclass{elsart}

\usepackage{dsfont}

\usepackage{amssymb}
\usepackage{amsmath}

\usepackage{indentfirst}
\usepackage{graphicx}
\usepackage{psfrag}

\usepackage[usenames,dvipsnames]{pstricks}
\usepackage{epsfig}
\usepackage{pst-grad} 
\usepackage{pst-plot} 

\usepackage{subfigure}

\usepackage[english]{babel}
\usepackage[latin1]{inputenc}

\usepackage{color}
\definecolor{oneblue}{rgb}{0,0.0,0.75}
\usepackage[colorlinks,
            urlcolor=oneblue,
            linkcolor=oneblue,
            citecolor=oneblue,
            bookmarksopen=false,
            pagebackref]{hyperref}

\journal{Studies in Applied Mathematics}

\newcommand{\grad}{\nabla}
\newcommand{\dt}{\partial_t}

\renewcommand{\div}{\nabla\cdot}
\newcommand{\od}[2]{\frac{d#1}{d#2}}

\newcommand{\tr}{\mathop{\mathrm{tr}}}
\newcommand{\area}{\mathop{\mathrm{area}}}
\newcommand{\vol}{\mathop{\mathrm{vol}}}
\newcommand{\Mat}{\mathop{\mathrm{Mat}}}
\newcommand{\set}[1]{\left\{ #1 \right\}}
\newcommand{\range}{\mathop{\mathrm{range}}}
\newcommand{\sign}{\mathop{\mathrm{sign}}}
\newcommand{\pd}[2]{\frac{\partial#1}{\partial#2}}

\newcommand{\F}{\mathcal{F}}
\newcommand{\g}{\vec{g}}
\newcommand{\q}{\vec{q}\,}
\newcommand{\x}{\vec{x}}
\newcommand{\n}{\vec{n}}

\newcommand{\ub}{\vec{u}}
\newcommand{\Ma}{\mathrm{Ma}}
\newcommand{\Fr}{\mathrm{Fr}}
\renewcommand{\Re}{\mathrm{Re}}
\newcommand{\St}{\mathrm{St\:}}
\newcommand{\T}{\mathbf{T}}
\newcommand{\B}{\mathbf{B}}
\newcommand{\A}{\mathbf{A}}
\newcommand{\R}{\mathbb{R}}
\newcommand{\M}{\mathbb{M}}
\newcommand{\Fd}{\vec{F}_d}
\newcommand{\D}{\mathbb{D}}
\newcommand{\K}{\mathbb{K}}
\newcommand{\J}{\mathbf{J}}
\newcommand{\Rr}{\mathbf{R}}
\newcommand{\Id}{\mathbf{I}}
\renewcommand{\u}{\vec{u}\,}
\renewcommand{\phi}{\varphi}
\newcommand{\Dv}{\mathbf{D}}
\renewcommand{\P}{\mathbf{P}}
\renewcommand{\S}{\mathbf{S}}
\newcommand{\Ve}{{V_\epsilon}}
\renewcommand{\O}{\mathcal{O}}
\renewcommand{\L}{\mathcal{L}}
\newcommand{\eps}{\varepsilon}
\newcommand{\ttau}{\boldsymbol{\tau}}
\newcommand{\w}{\mathbf{w}}

\newcommand{\Tt}{\mathcal{T}}
\newcommand{\N}{\mathcal{N}}

\begin{document}

\begin{frontmatter}

\title{Velocity and energy relaxation in two-phase flows}

\author[Y. Meyapin]{Yannick Meyapin}
\address[Y. Meyapin]{LAMA, UMR 5127 CNRS, Universit\'e de Savoie, 73376 Le Bourget-du-Lac Cedex, France}
\ead{Yannick.Meyapin@etu.univ-savoie.fr}

\author[D. Dutykh]{Denys Dutykh\corauthref{cor}}
\address[D. Dutykh]{LAMA, UMR 5127 CNRS, Universit\'e de Savoie, 73376 Le Bourget-du-Lac Cedex, France}
\corauth[cor]{Corresponding author.}
\ead{Denys.Dutykh@univ-savoie.fr}

\author[M. Gisclon]{Marguerite Gisclon}
\address[M. Gisclon]{LAMA, UMR 5127 CNRS, Universit\'e de Savoie, 73376 Le Bourget-du-Lac Cedex France}
\ead{Marguerite.Gisclon@univ-savoie.fr}

\begin{abstract}
In the present study we investigate analytically the process of velocity and energy relaxation in two-phase flows. We begin our exposition by considering the so-called six equations two-phase model \cite{Ishii1975,Rovarch2006}. This model assumes each phase to possess its own velocity and energy variables. Despite recent advances, the six equations model remains computationally expensive for many practical applications. Moreover, its advection operator may be non-hyperbolic which poses additional theoretical difficulties to construct robust numerical schemes \cite{Ghidaglia2001}. In order to simplify this system, we complete momentum and energy conservation equations by relaxation terms. When relaxation characteristic time tends to zero, velocities and energies are constrained to tend to common values for both phases. As a result, we obtain a simple two-phase model which was recently proposed for simulation of violent aerated flows \cite{Dias2008}. The preservation of invariant regions and incompressible limit of the simplified model are also discussed. Finally, several numerical results are presented.
\end{abstract}


\end{frontmatter}

\maketitle

\section{Introduction}

Currently, single phase flows modeling is essentially understood besides the problem of turbulence. In two-phase flows the situation is completely different. Nowadays there is no general consensus on the two-phase flow modeling. Turbulence modeling in two-phase flows is even more tricky \cite{Llor2005}. The two-phase models are based on space and time (ensemble) averaging of the local governing equations of each phase. Consequently, these models can only provide information on the average flow behaviour. The derivation of the \textit{proper} model is far from being achieved \cite{Ishii1975,Drew1979,Drew1979a,Boure1982,Stewart1984,Drew1994,Ishii2006}. One of the main difficulties lies in the determining the mass, momentum and energy transfers in the presence of steep gradients across the interfaces. Very often more or less accurate empirical correlations are used to describe such interface processes. In the same time, two-phase flows are very frequent in industry and in nature. Typical examples include water waves (especially during the wave breaking), petroleum and gas flow in pipes, nuclear reactors \cite{Staedtke2005}, etc.

The most important industrial application of two-phase flows is the simulation of a wide spectrum of accident scenarios in Pressurized (or Boiling) Water Reactors (PWR, BWR). Extensive experimental programmes for the PWRs are extremely expensive. Currently, new reactor design requires more and more numerical simulations of basic reactor features \cite{Staedtke2005}. The goal is to reduce experimental programmes to a minimum. On the other hand, industrial codes (such as CATHARE \cite{Bestion1990}, RELAP5 \cite{Ransom1985}, THYC \cite{Aubry1995} or Neptune\_CFD \cite{Bestion2005,Galassi2009,Guelfi2007}, for example) need to be supplemented by a large number of empirical closures which prevents reliable code application outside of their validity domain. Also these codes use robust but higly diffusive numerical schemes which make it difficult the computation of strong variable gradients proper to many accident conditions.

Two-phase models play also an important role in simulation of interfaces between two immiscible compressible fluids. This problem is very challenging from numerical point of view. All existing methods to address this problem can be conventionally divided into four big groups:
\begin{itemize}
  \item Volume of fluid (VOF) \cite{Hirt1981,Scardovelli1999}
  \item Level set method \cite{Osher1988,Osher1994}
  \item Lagrangian interface tracking \cite{Glimm2000}
  \item Diffuse interface method \cite{Larrouturou1990,Anderson1998,Shyue1998}
\end{itemize}
In the last approach we do not compute precisely the interface position. Due to numerical diffusion, the volume fraction takes intermediate values between 0 and 1 even if it is initialized by values in $\{0, 1\}$. The resulting mixture of two fluids has to be properly modelled in the transition region. Two-phase models that we will discuss below provide a sound physical description of mixing zones. The diffuse interface approach is less accurate than mentioned above methods (unless some special sharpening techniques are employed) but it is much simpler to implement and can naturally handle arbitrary topological changes in the flow.

Recently, a single velocity/single energy two-phase model was proposed \cite{Dutykh2007a} \cite{DDG2008,Dias2008a,Dias2008} in ad-hoc way as a model for violent aerated flows. By analogy with the six equations model, this system was called the four equations model since in one space dimension it consists of two equations of the mass conservation, one equation of the momentum and one of the energy conservation. Two-phase models where both fluids share the same velocity were previously considered in \cite{Allaire2002,Dellacherie2005}, for example.

The main goal of the present paper is to show the connection between the six equations model \cite{Ghidaglia2001,Ghidaglia2005,Rovarch2006,Ishii2006} (two equations of mass, momentum and energy conservation) and recently proposed four equations model \cite{Dutykh2007a,DDG2008,Dias2008a,Dias2008} with single velocity and single energy. Physically it is done by introducing relaxation terms into momentum and energy relaxation equations. These relaxation terms constrain velocities and energies to tend to a common value. Then, after taking the singular limit when the characteristic relaxation time goes to zero, one obtains a simplified model. Mathematically, it is achieved by employing a Chapman-Enskog type expansion \cite{Chapman1995}. This technique has been already successfully applied to the so-called Baer-Nunziato model \cite{Baer1986a,Baer1986} in \cite{Murrone2005}. The present study is greatly inspired by their work. This type of reduction from the barotropic six equations model were recently done in \cite{Meyapin2009a}.

The present manuscript is organized as follows. We start our exposition by presenting the so-called six equations model in Section \ref{sec:model}. Then, we complete this model by relaxation terms and derive in Section \ref{sec:relax} a single velocity, single energy model. Preservation of invariant regions and incompressible limit of the resulting four equations model are studied in Sections \ref{sec:inv} and \ref{sec:Mach} correspondingly. Then we present some numerical results in Section \ref{sec:num}. Finally, this paper is ended by drawing out main conclusions of this study and some perspectives for future research (Section \ref{sec:concl}).


\section{Mathematical model}\label{sec:model}

\begin{figure}
	\centering
		\includegraphics[width=0.80\textwidth]{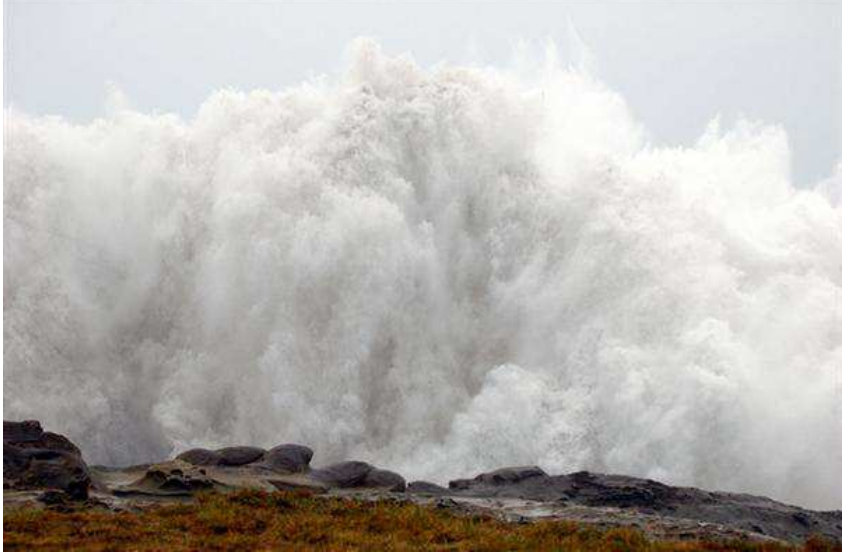}
	\caption{A quasi uniform air/water mixture after the wave breaking has occured.}
	\label{fig:BreakingWaves}
\end{figure}

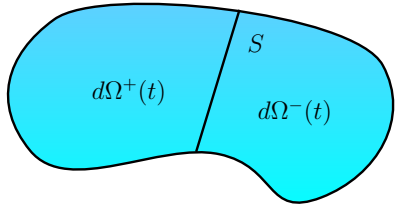
\begin{figure}
  \begin{center}
	\scalebox{0.8} 
	{
	\begin{pspicture}(0,-1.9729664)(6.9945483,1.9723125)
	\definecolor{color1g}{rgb}{0.4,0.8,1.0}
	\psbezier[linewidth=0.04,fillstyle=gradient,gradlines=2000,gradbegin=color1g,gradmidpoint=1.0](3.7758446,-0.87319964)(4.7539034,-1.0815277)(4.5428023,-1.9529666)(5.4758444,-1.5931996)(6.408887,-1.2334328)(6.9745483,-0.27117625)(6.4958444,0.6068004)(6.017141,1.484777)(1.8465083,1.9523125)(1.0358446,1.3668003)(0.22518091,0.7812883)(0.0030853755,-0.038851038)(0.6358446,-0.81319964)(1.2686038,-1.5875483)(2.7977855,-0.66487163)(3.7758446,-0.87319964)
	\psline[linewidth=0.04cm](3.3958447,-0.8331996)(4.1158447,1.5268004)
	\usefont{T1}{ptm}{m}{n}
	\rput(2.2772508,0.15680036){$d\Omega^+ (t)$}
	\usefont{T1}{ptm}{m}{n}
	\rput(5.037251,-0.18319964){$d\Omega^- (t)$}
	\usefont{T1}{ptm}{m}{n}
	\rput(4.387251,0.9768004){$S$}
	\end{pspicture}
	}
	\caption{An elementary fluid volume $d\Omega$ occupied by two phases.}
	\label{fig:volumefract}
	\end{center}
\end{figure}

Consider a fluid domain $\Omega \subseteq \R^3$ which is filled by two miscible fluids. All physical quantities related to the heavy and light fluids will be denoted by the superscripts $+$ and $-$ correspondingly. For example, if we consider a plunging breaker with important mixing processes, then water velocity, density, viscosity, etc. will be denoted with the superscript $+$, while variables related to the air will have the superscript $-$. When the interfaces are so multiple that it is impossible any more to follow them (see Figure \ref{fig:BreakingWaves} for an illustration), the classical modelling procedure consists in applying a volume average operator \cite{Ishii1975,Rovarch2006}. At this level of description, two additional variables naturally appear. The so-called volume fractions $\alpha^\pm (\x, t)$, $\x\in\Omega$ are defined as
\begin{equation}\label{eq:alphaDef}
  \alpha^\pm(\x,t) := \lim\limits_{\stackrel{|d\Omega|\to 0}{\x \in d\Omega}}
  \frac{|d\Omega^\pm|}{|d\Omega|},
\end{equation}
the heavy fluid occupies volume $d\Omega^+\subseteq d\Omega$ and the light one the volume $d\Omega^-\subseteq d\Omega$ (see Figure \ref{fig:volumefract}) such that
\begin{equation}\label{eq:dOmega}
  |d\Omega| = |d\Omega^+| + |d\Omega^-|.
\end{equation}
After taking the limit in relation (\ref{eq:dOmega}), one readily finds
\begin{equation}\label{eq:alpha}
  \alpha^+(\x,t) + \alpha^-(\x,t) \equiv 1, \quad \forall (\x,t) \in \Omega\times[0, T].
\end{equation}
The volume fractions $\alpha^\pm$ characterize the volume occupied by the corresponding phase per unit volume of the mixture.

After applying the averaging procedure, we obtain the so-called six equations model \cite{Ishii1975,Stewart1984,Ghidaglia2001,Rovarch2006}:
\begin{equation}\label{eq:mass6}
  \dt (\alpha^{\pm}\rho^{\pm}) + \div(\alpha^{\pm}\rho^{\pm}\u^{\pm}) = 0,
\end{equation}
\begin{equation}\label{eq:mom6}
  \dt (\alpha^{\pm}\rho^{\pm}\u^{\pm}) + \div(\alpha^{\pm}\rho^{\pm}\u^{\pm}\otimes\u^{\pm}) +
  \alpha^{\pm}\nabla p = \div(\alpha^{\pm}\ttau^{\pm}) + \alpha^{\pm}\rho^{\pm}\g,
\end{equation}
\begin{multline}\label{eq:energy6}
  \dt (\alpha^\pm\rho^\pm E^\pm) + \div(\alpha^\pm\rho^\pm H^\pm\u^\pm) + p\:\dt\alpha^{\pm}
  = \\ 
  \div(\alpha^\pm\ttau^\pm\u^\pm) - \div(\alpha^\pm\q^\pm) + \alpha^\pm\rho^\pm\g\cdot\u^\pm,
\end{multline}
where $\rho^\pm (\x,t)$, $\u^\pm (\x,t)$, $E^\pm (\x,t)$ are densities, velocities and energies of each fluid respectively. The total energy $E^\pm$ is the sum of the internal and kinetic energies $E^\pm := e^\pm + \frac12|\u^\pm|^2$. The specific total enthalpy $H^\pm$ is defined as $H^\pm := h^\pm + \frac12|\u^\pm|^2 = E^\pm + \displaystyle\frac{p}{\rho^\pm}$, where $h^\pm$ is just the specific enthalpy. Finally, $\g$ stands for the vector of the gravity acceleration.

The symbol $\ttau^\pm$ denotes the viscous stress tensor of each phase. If we assume both fluids to be Newtonian, the viscous stress tensor $\ttau^\pm$ can be written as
\begin{equation}\label{eq:viscous}
  \ttau^\pm = \lambda^{\pm}\tr\D(\u^{\pm})\Id + 2\mu^{\pm} \D(\u^{\pm}), \quad
  \tr\D(\u^{\pm}) = \div\u^\pm,
\end{equation}
where $\Id := (\delta_{ij})_{1\leq i,j\leq 3}$ is the identity tensor, $\D(\u) := \frac12\Bigl(\nabla\u + {}^t(\nabla\u)\Bigr)$ is the deformation rate and $\lambda^\pm$, $\mu^\pm$ are viscosity coefficients. For ideal gases, for example, these coefficients are related by Stokes relation $\lambda^\pm + \frac23\mu^\pm = 0$. Recall that the derivation procedure presented in the next section does not require any particular form of the tensor $\ttau^\pm$.

Finally, the heat flux in each fluid is denoted by $\q^\pm$ (see Equation (\ref{eq:energy6})). As with viscous stress tensor, we do not assume any particular form of the heat flux. However, in most cases the Fourier's law \cite{Fourier1822} is adopted:
\begin{equation*}
  \q^\pm := -K^\pm\grad T^\pm,
\end{equation*}
where $K^\pm$ is the thermal conductivity coefficient and $T^\pm$ is the thermodynamic temperature.

\begin{rem}
When a material demonstrates flagrant anisotropic properties, it is advised to use the generalized Fourier's law \cite{Ishii2006}:
\begin{equation*}
  \q^\pm := -\K^\pm\cdot\grad T^\pm,
\end{equation*}
where $\K^\pm$ is the conductivity tensor. The same remark applies to the viscous stress tensor $\ttau^\pm$ as well.
\end{rem}

In order to close the system (\ref{eq:mass6}) -- (\ref{eq:energy6}), we have to provide two equations of state which relate the pressure $p$ to other thermodynamic variables $p = p^\pm(\rho^\pm, e^\pm)$. We assume the equations of state to satisfy some general thermodynamic assumptions:
\begin{equation}\label{eq:EOS}
  p^\pm(\rho^\pm, e^\pm) > 0, \quad \left.\pd{p^\pm (\rho^\pm, e^\pm)}{\rho^\pm}\right|_{s^\pm} > 0, \quad \mbox{for} \quad \rho^\pm > 0.
\end{equation}
The symbol $\left.\pd{p^\pm (\rho^\pm, e^\pm)}{\rho^\pm}\right|_{s^\pm}$ denotes the partial differentiation with respect to the density $\rho^\pm$ while the entropy $s^\pm$ is kept constant. Physically this quantity corresponds to the squared velocity of pressure waves in each fluid (see Lemma \ref{lem:Gibbs}).

\begin{rem}
We reiterate that the model (\ref{eq:mass6}) -- (\ref{eq:energy6}) is a \textit{one pressure} model, i.e. two phases share the same pressure. Models with two pressures can also be derived \cite{Stewart1984,Baer1986,Baer1986a,Ransom1988,Murrone2005}. In this situation, an extra closure (algebraic or differential) must be added. In many cases this additional closure represents a relaxation process which will tend to equilibrate the pressures.
\end{rem}

\begin{rem}
The momentum balance Equation (\ref{eq:mom6}) can be rewritten differently:
\begin{equation}\label{eq:mom6bis}
  \dt (\alpha^{\pm}\rho^{\pm}\u^{\pm}) + \div\bigl(\alpha^{\pm}\rho^{\pm}\u^{\pm}\otimes\u^{\pm} + \alpha^\pm p\Id\bigr)
  -p\nabla\alpha^{\pm} = \div(\alpha^{\pm}\ttau^{\pm}) + \alpha^{\pm}\rho^{\pm}\g.
\end{equation}
These two forms are obviously equivalent for smooth solutions. However, it is not the case for discontinuous ones. We believe that this form is more relevant from physical point of view, since the flux of momentum involves the pressure. In our analytical investigations we consider smooth solutions, however for numerics we advise using the last form (\ref{eq:mom6bis}).
\end{rem}

\begin{rem}\label{rem:massf}
While considering two-phase flows, it is useful to introduce several additional quantities which play an important r\^ole in the description of such flows. The mixture density $\rho$ and mass fractions $m^\pm$ are naturally defined as
\begin{equation}\label{eq:mixtdens}
  \rho (\x,t) := \alpha^+\rho^+ + \alpha^-\rho^- > 0, \quad \forall (\x,t) \in \Omega\times [0, T], 
\end{equation}
\begin{equation*}
  m^\pm := \frac{\alpha^\pm\rho^\pm}{\rho}, \quad \mbox{consequently} \quad m^+ + m^- = 1.
\end{equation*}
The total density $\rho$ is assumed to be strictly positive everywhere in the domain $\Omega$. Hence, the void creation is forbidden in our modeling paradigm. Important quantities $\rho$, $m^\pm$ will appear several times below.
\end{rem}

The six-equations two-phase model presented in this section contains some other simplifications. Namely, we neglect capillarity effects which could be taken into account in the form of the Korteweg term \cite{Korteweg1901,Bresch2008}. Moreover, terms modeling mass, momentum and energy exchange between the phases are neglected as well. In general, their form depends strongly on the flow regime under consideration and is a subject of current debates.

\begin{rem}
Even if the Jacobian matrix of the advection operator in (\ref{eq:mass6}) -- (\ref{eq:energy6}) is not diagonalizable over $\R$, the existence of global weak solutions was recently proven at least in barotropic case \cite{Bresch2008} if we supplement the model by viscous and surface tension effects \cite{Korteweg1901}. Previously, it was already shown that viscosity and thermal diffusivity are indeed sufficient for the evolution problem well-posedness in the linearized or in the nonlinear sense for small data \cite{Arai1980,Ramos2000}.
\end{rem}

Despite recent advances in numerical methods \cite{Ghidaglia1996,Ghidaglia2001,Rovarch2006} and several simplifications made above, the system (\ref{eq:mass6}) -- (\ref{eq:energy6}) is still very complex to solve numerically for industrial scale problems. One of the main difficulties lies in the advection operator which is known to be generally non-hyperbolic. In the next section we will derive a simplified two-phase model which has fewer variables and possesses the requested hyperbolic structure for quite general equations of state.

\section{Relaxation process}\label{sec:relax}

In this section we will supplement momentum and energy conservation Equations (\ref{eq:mom6}), (\ref{eq:energy6}) by relaxation terms. Physically, they represent the friction or drag force between two phases. When time evolves, this mechanism will ensure the convergence of two velocities $\u^\pm$ and energies $E^\pm$ to common for both phases values $\u$ and $E$. Augmented Equations (\ref{eq:mom6}), (\ref{eq:energy6}) take the following form:
\begin{equation}\label{eq:mom6d}
  \dt (\alpha^{\pm}\rho^{\pm}\u^{\pm}) + \div(\alpha^{\pm}\rho^{\pm}\u^{\pm}\otimes\u^{\pm}) +
  \alpha^{\pm}\nabla p = \div(\alpha^{\pm}\ttau^{\pm}) + \alpha^{\pm}\rho^{\pm}\g \pm \Fd,
\end{equation}
\begin{multline}\label{eq:energy6d}
  \dt (\alpha^\pm\rho^\pm E^\pm) + \div(\alpha^\pm\rho^\pm H^\pm\u^\pm) + p\:\dt\alpha^{\pm}
  = \\ 
  \div(\alpha^\pm\ttau^\pm\u^\pm) - \div(\alpha^\pm\q^\pm) + \alpha^\pm\rho^\pm\g\cdot\u^\pm \pm E_d.
\end{multline}
In the present study we choose friction terms $\Fd$ and $E_d$ in this form:
\begin{equation*}
  \Fd := \frac{\kappa}{\eps}\frac{\alpha^-\rho^-\alpha^+\rho^+}{\alpha^+\rho^+ + \alpha^-\rho^-} (\u^+ - \u^-), \quad
  E_d := \frac{\kappa}{\eps}\frac{\alpha^-\rho^-\alpha^+\rho^+}{\alpha^+\rho^+ + \alpha^-\rho^-}
  \ub\cdot(\u^+ - \u^-),
\end{equation*}
where $\ub := m^+\u^+ + m^-\u^-$ is the barycentric velocity, $\kappa = \O(1)$ is a dimensionless constant and $\eps$ is a small parameter which controls the magnitude of the relaxation term. Physically it represents the characteristic relaxation time. In the following we are going to take the singular limit as the relaxation parameter $\eps\to 0$. This goal is achieved with a Chapman-Enskog type expansion \cite{Chapman1995}.

This kind of computations has already been successfully carried out for the Baer-Nunziato model \cite{Baer1986}. In this section we follow in great lines the work of Guillard \& Murrone \cite{Murrone2005}. However, the computation details are substantially different.

The first step consists in rewriting the governing Equations (\ref{eq:mass6}), (\ref{eq:mom6d}), (\ref{eq:energy6d}) in quasilinear form. To shorten the notation, we will also use the material derivative which is classically defined for any smooth scalar function $\phi (\x,t)$ as
\begin{equation*}
  \od{^\pm\phi}{t} := \pd{\phi}{t} + \u^\pm\cdot\grad\phi.
\end{equation*}

In computations below, we will need the following technical lemma:
\begin{lem}\label{lem:Gibbs}
Consider two compressible fluids with general equations of state $p = p^\pm (\rho^\pm, e^\pm)$. Then, partial derivatives of pressure $p$ are given by:
\begin{equation*}
  \left.\pd{p}{s^\pm}\right|_{\rho^\pm} = \frac{T^\pm}{\theta^\pm}, \qquad
  \left.\pd{p}{\rho^\pm}\right|_{s^\pm} \equiv (c_s^\pm)^2 =
  \frac{1}{\theta^\pm}\Bigl(\frac{p}{(\rho^\pm)^2} - \chi^\pm\Bigr),
\end{equation*}
where $s^\pm$, $c_s^\pm$ are the entropy and the sound velocity in the phase $\pm$ respectively. We also denote by $\chi^\pm := \left.\pd{e^\pm}{\rho^\pm}\right|_{p}$ and by $\theta^\pm := \left.\pd{e^\pm}{p}\right|_{\rho^\pm}$.
\end{lem}

\begin{pf}
Obviously, one can write:
\begin{equation}\label{eq:dp}
  dp = \left.\pd{p}{\rho^\pm}\right|_{s^\pm} d\rho^\pm + \left.\pd{p}{s^\pm}\right|_{\rho^\pm} ds^\pm.
\end{equation}
Similarly, if one takes into account the definitions of $\chi^\pm$ and $\theta^\pm$:
\begin{equation}\label{eq:de}
  de^\pm = \chi^\pm d\rho^\pm + \theta^\pm dp.
\end{equation}
Now we will use the Gibbs relation \cite{Ishii1975,Callen1985,Ishii2006}:
\begin{equation}\label{eq:Gibbs}
  de^\pm = T^\pm ds^\pm + \frac{p}{(\rho^\pm)^2}d\rho^\pm.
\end{equation}
After expanding $de^\pm$ and $dp$ according to (\ref{eq:de}), (\ref{eq:dp}) correspondingly, we get the following differential inequality:
\begin{equation*}
  \Bigl(\frac{p}{(\rho^\pm)^2} - \chi^\pm -
  \theta^\pm\left.\pd{p}{\rho^\pm}\right|_{s^\pm}\Bigr)d\rho^\pm
  + \Bigl(T^\pm-\theta^\pm \left.\pd{p}{s^\pm}\right|_{\rho^\pm}\Bigr)ds^\pm = 0.
\end{equation*}
Since this equality must be true for arbitrary $de^\pm$ and $dp$, we conclude the proof by requiring the coefficients to be equal to zero.
\end{pf}

\begin{thm}\label{thm:noncons}
Smooth solutions to Equations (\ref{eq:mass6}), (\ref{eq:mom6d}), (\ref{eq:energy6d}) satisfy the following system:
\begin{equation}\label{eq:entropy}
  \alpha^\pm\rho^\pm T^\pm \od{^\pm s^\pm}{t} = \alpha^\pm\ttau^\pm:\grad\u^\pm - \div(\alpha^\pm\q^\pm) \pm (E_d - \Fd\cdot\u^\pm),
\end{equation}
\begin{equation}\label{eq:mom6drag}
  \alpha^\pm\rho^\pm\od{^\pm\u^\pm}{t} + \alpha^\pm\grad p = \div(\alpha^\pm\ttau^\pm) + \alpha^\pm\rho^\pm\g \pm \Fd,
\end{equation}
\begin{multline}\label{eq:p}
  \alpha^\pm\rho^\pm\theta^\pm\od{^\pm p}{t} \pm (\rho^\pm)^2(c_s^\pm)^2\theta^\pm (\dt\alpha^+ \pm \div(\alpha^\pm\u^\pm)) = \\ \alpha^\pm\ttau^\pm:\grad\u^\pm - \div(\alpha^\pm\q^\pm) \pm (E_d - \Fd\cdot\u^\pm),
\end{multline}
where $s^\pm$ is the entropy, $(c_s^\pm)^2$ and $\theta^\pm$ are defined in Lemma \ref{lem:Gibbs}.
\end{thm}

\begin{pf}
The mass conservation Equation (\ref{eq:mass6}) is straightforwardly rewritten using the material derivative:
\begin{equation}\label{eq:massLagr}
  \od{^\pm(\alpha^\pm\rho^\pm)}{t} + \alpha^\pm\rho^\pm\div\u^\pm = 0.
\end{equation}
If we multiply the latter by $\u^\pm$ and subtract the result from (\ref{eq:mom6d}), we will obtain announced above Equation (\ref{eq:mom6drag}). Similarly, multiplying Equation (\ref{eq:massLagr}) by $E^\pm$ and subtracting it from (\ref{eq:energy6d}) leads to the total energy equation in quasilinear form:
\begin{multline}\label{eq:totalE}
  \alpha^\pm\rho^\pm\od{^\pm E^\pm}{t} + \div(\alpha^\pm p\u^\pm) + p\alpha^\pm_t = \\
  \div(\alpha^\pm\ttau^\pm\u^\pm) - \div(\alpha^\pm\q^\pm) + \alpha^\pm\rho^\pm\g\cdot\u^\pm \pm E_d.
\end{multline}
The kinetic energy evolution equation is straightforwardly obtained after taking the scalar product of (\ref{eq:mom6drag}) with $\u^\pm$:
\begin{equation}\label{eq:kineticE}
  \alpha^\pm\rho^\pm\od{^\pm(\frac12|\u^\pm|^2)}{t} + \alpha^\pm\u^\pm\cdot\grad p = \u^\pm\cdot\div(\alpha^\pm\ttau^\pm) + \alpha^\pm\rho^\pm\g\cdot\u^\pm \pm \Fd\cdot\u^\pm.
\end{equation}
Recall that the total energy $E^\pm$ is constituted of internal and kinetic energies: $E^\pm = e^\pm + \frac12|\u^\pm|^2$. Consequently, the internal energy evolution equation is derived by subtracting (\ref{eq:kineticE}) from (\ref{eq:totalE}):
\begin{equation}\label{eq:internalE}
  \alpha^\pm\rho^\pm\od{^\pm e^\pm}{t} + p\div(\alpha^\pm\u^\pm) + p\alpha^\pm_t = \alpha^\pm\ttau^\pm:\grad\u^\pm - \div(\alpha^\pm\q^\pm) \pm (E_d - \Fd\cdot\u^\pm).
\end{equation}
In order to introduce the entropy variable $s^\pm$ into our considerations, we will make use of the Gibbs relation. After multiplying (\ref{eq:Gibbs}) by $\alpha^\pm\rho^\pm$ and passing to the material derivative, one gets:
\begin{equation*}
  \alpha^\pm\rho^\pm T^\pm\od{^\pm s^\pm}{t} = \alpha^\pm\rho^\pm\od{^\pm e^\pm}{t} 
  - \frac{\alpha^\pm p}{\rho^\pm}\od{^\pm\rho^\pm}{t}.
\end{equation*}
Substituting (\ref{eq:internalE}) into the last relation and taking into account (\ref{eq:massLagr}), leads to the requested entropy Equation (\ref{eq:entropy}).

From lemma \ref{lem:Gibbs} we directly obtain the following equality:
\begin{equation*}
  \alpha^\pm\rho^\pm\theta^\pm\od{^\pm p}{t} = 
  \alpha^\pm\rho^\pm\Bigl(\frac{p}{(\rho^\pm)^2} - \chi^\pm\Bigr)\od{^\pm\rho^\pm}{t} 
  + \alpha^\pm\rho^\pm T^\pm\od{^\pm s^\pm}{t}.
\end{equation*}
After substituting (\ref{eq:entropy}) and replacing $\bigl(\frac{p}{(\rho^\pm)^2} - \chi^\pm\bigr)$ by  $\theta^\pm(c_s^\pm)^2$, we obtain (\ref{eq:p}) and the proof is completed.
\end{pf}

Computations that we will perform below will be clearer if the governing Equations (\ref{eq:entropy}) -- (\ref{eq:p}) are recast in the vectorial form:
\begin{equation}\label{eq:matrix}
  \A(\Ve)\pd{\Ve}{t} + \B(\Ve)\nabla\Ve = \div\T(\Ve) + \S(\Ve) + \frac{\Rr(\Ve)}{\epsilon},
\end{equation}
where we introduced several notations. The vector $\Ve$ represents six unknown physical variables $\Ve := {}^t (s^+, s^-, \u^+, \u^-, p, \alpha^+)$ and $\pd{\Ve}{t}$ is the componentwise partial time derivative. Symbol $\nabla\Ve$ denotes this vector:
\begin{equation*}
\nabla\Ve := {}^t\bigl(\grad s^+, \grad s^-, (\cdot\nabla)\u^+, (\cdot\nabla)\u^-, \grad p, \grad\alpha^+\bigr).
\end{equation*}
Matrices $\A(\Ve)$ and $\B(\Ve)$ are defined as
\begin{equation*}
  \A(\Ve) := \begin{pmatrix}
  						a_T^+ & 0 & 0 & 0 & 0 & 0 \\
  						0 & a_T^- & 0 & 0 & 0 & 0 \\
  						0 & 0 & \alpha^+\rho^+\Id & 0 & 0 & 0 \\
  						0 & 0 & 0 & \alpha^-\rho^-\Id & 0 & 0 \\
  						0 & 0 & 0 & 0 & a_\theta^+ & p_\theta^+ \\
  						0 & 0 & 0 & 0 & a_\theta^- & -p_\theta^- \\
  					 \end{pmatrix}
\end{equation*}
\begin{equation*}
  \B(\Ve) := \begin{pmatrix}
  						a_T^+\u^+ & 0 & 0 & 0 & 0 & 0 \\
  						0 & a_T^-\u^- & 0 & 0 & 0 & 0 \\
  						0 & 0 & \alpha^+\rho^+\u^+ & 0 & \alpha^+\Id & 0 \\
  						0 & 0 & 0 & \alpha^-\rho^-\u^- & \alpha^-\Id & 0 \\
  						0 & 0 & \alpha^+p_\theta^+\Id & 0 & a_\theta^+\u^+ & p_\theta^+\u^+ \\
  						0 & 0 & 0 & \alpha^-p_\theta^-\Id & a_\theta^-\u^- & -p_\theta^-\u^- \\
  					 \end{pmatrix},
\end{equation*}
where several symbols were introduced to shorten the notation:
\begin{equation*}
  p_\theta^\pm := (\rho^\pm)^2(c_s^\pm)^2\theta^\pm, \quad
  a_T^\pm := \alpha^\pm\rho^\pm T^\pm, \quad
  a_\theta^\pm := \alpha^\pm\rho^\pm\theta^\pm.
\end{equation*}
In these matrix notations the size of zero entries must be chosen to make the multiplication operation possible.

On the right hand side of (\ref{eq:matrix}), the work of viscous forces is denoted by symbol $\div\T(\Ve)$ which is defined as
\begin{multline*}
 \div\T(\Ve) := {}^t(\alpha^+\ttau^+:\grad\u^+ - \div(\alpha^+\q^+), \alpha^-\ttau^-:\grad\u^- - \div(\alpha^-\q^-), \\ \div(\alpha^+\ttau^+), \div(\alpha^-\ttau^-), \alpha^+\ttau^+:\grad\u^+ - \div(\alpha^+\q^+), \alpha^-\ttau^-:\grad\u^- - \div(\alpha^-\q^-)).
\end{multline*}
The source term $\S(\Ve) := {}^t(0, 0, \alpha^+\rho^+\g, \alpha^-\rho^-\g, 0, 0)$ incorporates the gravity force and $\Rr(\Ve)$ contains the relaxation terms:
\begin{multline*}
  \Rr(\Ve) := \kappa\frac{\alpha^+\rho^+\alpha^-\rho^-}{\alpha^+\rho^+ + \alpha^-\rho^-} ^{t}\Bigl(\ub - \u^+, -(\ub - \u^-), \\
  1, -1, \ub - \u^+, -(\ub - \u^-)\Bigr)\times(\u^+ - \u^-).
\end{multline*}
Since we expect the limit $\Ve\to V$ to be finite as $\epsilon\to 0$, necessary the limiting vector $V$ lies in the hypersurface $\Rr(V) = 0$. In terms of physical variables, it implies $\u = \u^+ = \u^-$. Consequently, we find our solution in the form of the following Chapman-Enskog type expansion \cite{Chapman1995}:
\begin{equation*}
  \Ve = V + \eps W + \O(\eps^2).
\end{equation*}
After substituting this expansion into (\ref{eq:matrix}) and taking into account the fact that $\Rr(V) \equiv 0$, at the leading order in $\epsilon$ one obtains:
\begin{equation}\label{eq:asympt}
  \A(V)\pd{V}{t} + \B(V)\nabla V = \div\T(V) + \S(V) + \Rr'(V)W,
\end{equation}
where 
\begin{equation*}
  \Rr'(V) := \kappa\frac{\alpha^+\rho^+\alpha^-\rho^-}{\alpha^+\rho^+ + \alpha^-\rho^-}
  					\begin{pmatrix}
   					  0 & 0 & 0 & 0 & 0 & 0 \\
   					  0 & 0 & 0 & 0 & 0 & 0 \\
   					  0 & 0 & \Id & -\Id & 0 & 0 \\
   					  0 & 0 & -\Id & \Id & 0 & 0 \\
   					  0 & 0 & 0 & 0 & 0 & 0 \\
   					  0 & 0 & 0 & 0 & 0 & 0 \\   					  
  					 \end{pmatrix}
\end{equation*}

Henceforth, we make a technical assumption of the presence of both phases in any point $\x\in\Omega$ of the flow domain. Mathematically it means that $0 < \alpha^+ < 1$. Since $\alpha^+ + \alpha^- = 1$, obviously the same inequality holds for $\alpha^-$. Otherwise, the relaxation process physically does not make sense and we will have some mathematical technical difficulties.

Under the aforementioned assumption, the matrix $\A(V)$ is invertible. Hence, we can multiply on the left both sides of (\ref{eq:asympt}) by $\P\A^{-1}(V)$ where the projection matrix $\P$ is to be specified below:
\begin{equation}\label{eq:Peq}
  \P\pd{V}{t} + \P\A^{-1}(V)\B(V)\nabla V = \P\A^{-1}(V)\div\T(V) + \P\tilde\Rr'(V)W + \P\A^{-1}(V)\S(V),
\end{equation}
where $\tilde\Rr'(V) := \A^{-1}(V)\Rr'(V)$ and has the following components
\begin{equation*}
  \tilde\Rr'(V) = \kappa\frac{\alpha^+\rho^+\alpha^-\rho^-}{\alpha^+\rho^+ + \alpha^-\rho^-}
  					\begin{pmatrix}
   					  0 & 0 & 0 & 0 & 0 & 0 \\
   					  0 & 0 & 0 & 0 & 0 & 0 \\
   					  0 & 0 & \displaystyle\frac{1}{\alpha^+\rho^+}\Id & -\displaystyle\frac{1}{\alpha^+\rho^+}\Id & 0 & 0 \\
   					  0 & 0 & -\displaystyle\frac{1}{\alpha^-\rho^-}\Id & \displaystyle\frac{1}{\alpha^-\rho^-}\Id & 0 & 0 \\
   					  0 & 0 & 0 & 0 & 0 & 0 \\
   					  0 & 0 & 0 & 0 & 0 & 0 \\   					  
  					 \end{pmatrix}
\end{equation*}

The vector of physical variables $V$ has six components (in 1D):
\begin{equation*}
  V = {}^t(s^+, s^-, \u, \u, p, \alpha^+)
\end{equation*} 
and only five are different. In order to remove the redundant information, we will introduce the new vector $U$ defined as $U := {}^t(s^+, s^-, \u, p, \alpha^+)$. The Jacobian matrix of this transformation can be easily computed:
\begin{equation*}
  \J := \pd{V}{U} = \begin{pmatrix}
  										1 & 0 & 0 & 0 & 0 \\
  										0 & 1 & 0 & 0 & 0 \\
  										0 & 0 & \Id & 0 & 0 \\
  										0 & 0 & \Id & 0 & 0 \\
  										0 & 0 & 0 & 1 & 0 \\
  										0 & 0 & 0 & 0 & 1 \\
  									\end{pmatrix}.
\end{equation*}
In new variables Equation (\ref{eq:Peq}) becomes:
\begin{equation}\label{eq:Peq2}
	\P\J\pd{U}{t} + \P\A^{-1}(U)\B(U)\J\nabla U = \P\A^{-1}(U)\div\T(U) + \P\tilde\Rr'(U)W + \P\A^{-1}(U)\S(U).
\end{equation}

Now we can formulate two conditions to construct the matrix $\P$. First of all, the vector $W$ is unknown and we need to remove it from Equation (\ref{eq:Peq2}). Hence, we require $\P\tilde\Rr'(V) = 0$. Then, we would like the governing equations to be explicitly resolved with respect to time derivatives. It gives us the second condition $\P\J = \Id$. The existence and effective construction of the matrix $\P$ satisfying two aforementioned conditions
\begin{equation*}
  \P\tilde\Rr'(V) = 0, \quad \P\J = \Id,
\end{equation*}
are discussed below. Presented in this section results follow in great lines \cite{Murrone2005}.

We will consider a slightly more general setting. Let be vector $V\in\R^n$ and its reduced counterpart $U\in\R^{n-k}$, $k < n$. In such geometry, $\tilde\Rr'(V)\in\Mat_{n,n}(\R)$, $\J\in\Mat_{n,n-k}(\R)$ and, consequently, $\P\in\Mat_{n-k,n} (\R)$. Here, the notation $\Mat_{m,n}(\R)$ denotes the set of $m \times n$ matrices with coefficients in $\R$. We have to say also that from algebraic point of view, matrices $\tilde\Rr'(V)$ and $\Rr'(V)$ are completely equivalent. Thus, for the sake of simplicity, the following propositions will be formulated in terms of the matrix $\Rr'(V)$.

\begin{lem}\label{lemma:J}
The columns of the Jacobian matrix $\J$ form a basis of $\ker\bigl(\Rr'(V)\bigr)$.
\end{lem}

\begin{pf}
If we differentiate the relation $\Rr(V) = 0$ with respect to $U$, we will get the identity $\Rr'(V)\J = 0$. It implies that $\range\bigl(\J\bigr) \subseteq \ker \bigl(\Rr'(V)\bigr)$. By direct computation one verifies that $\dim\range\bigl(\Rr'(V)\bigr) = k$. From the well-known identity $\range\bigl(\Rr'(V)\bigr)\oplus \ker\bigl(\Rr'(V)\bigr) = \R^n$, one concludes that $\dim\ker\bigl(\Rr'(V)\bigr) = n-k$. But in the same time, the rank of $\J$ is equal to $n-k$ as well. It proves the result.
\end{pf}

\begin{thm}
We suppose that for all $V$, $\range\bigl(\Rr'(V)\bigr) \cap \ker\bigl(\Rr'(V)\bigr) = \{0\}$ then there exists a matrix $\P\in\Mat_{n-k,n} (\R)$ such that $\P\Rr'(V) = 0$ and $\P\J = \Id_{n-k}$.
\end{thm}

\begin{pf}
Assumption $\range\bigl(\Rr'(V)\bigr) \cap \ker\bigl(\Rr'(V)\bigr) = \{0\}$ implies that
\begin{equation*}
  \range\bigl(\Rr'(V)\bigr)\oplus \ker\bigl(\Rr'(V)\bigr) = \R^n.
\end{equation*}
From Lemma \ref{lemma:J} it follows that $\range\bigl(\J\bigr) = \ker\bigl(\Rr'(V)\bigr)$. Thus, the space $\R^n$ can be also represented as a direct sum $\range\bigl(\Rr'(V)\bigr)\oplus \range\bigl(\J\bigr)$. We will define $\P$ to be the projection on $\ker\bigl(\Rr'(V)\bigr) \equiv \range\bigl(\J\bigr)$. Since obviously $\Rr'(V)\in \range\bigl(\Rr'(V)\bigr)$ and $\J\in\range\bigl(\J\bigr)$, we have two required identities: $\P\J = \Id_{n-k}$ and $\P\Rr'(V) = 0$.
\end{pf}

Now, in order to compute effectively the projection matrix $\P$, we will construct an auxiliary matrix $\Dv(V) = [J^1,\ldots, J^{n-k}, I^1,\ldots,I^k]$, where $J^i$ is the column $i$ of the matrix $\J$ and $\{I^1,\ldots,I^k\}$ are vectors which form a basis of $\range\bigl(\Rr'(V)\bigr)$. We remark that $\P\Dv(V) = [\Id_{n-k}, 0]$. Lemma \ref{lemma:J} implies that the matrix $\Dv(V)$ is invertible. Thus, the projection $\P$ can be computed by inverting $\Dv(V)$:
\begin{equation}\label{eq:pcomput}
  \P = [\Id_{n-k}, 0]\cdot\Dv^{-1}(V).
\end{equation}

Let us apply this general framework to our model (\ref{eq:Peq}), where $n = 6$ and $k = 1$. The matrix $\Dv(V)$ and its inverse $\Dv^{-1}(V)$ take this form:
\begin{equation*}
  \Dv(V) = \begin{pmatrix}
  										1 & 0 & 0 & 0 & 0 & 0 \\
  										0 & 1 & 0 & 0 & 0 & 0 \\
  										0 & 0 & \Id & 0 & 0 &  m^-\Id \\
  										0 & 0 & \Id & 0 & 0 & -m^+\Id \\
  										0 & 0 & 0 & 1 & 0 & 0 \\
  										0 & 0 & 0 & 0 & 1 & 0 \\
  									\end{pmatrix}, \quad
\Dv^{-1}(V) = \begin{pmatrix}
  										1 & 0 & 0 & 0 & 0 & 0 \\
  										0 & 1 & 0 & 0 & 0 & 0 \\
  										0 & 0 & m^+\Id & m^-\Id & 0 & 0 \\
  										0 & 0 & 0 & 0 & 1 & 0 \\
  										0 & 0 & 0 & 0 & 0 & 1 \\
  										0 & 0 & \Id & -\Id & 0 & 0 \\
  									\end{pmatrix},
\end{equation*}
where $m^\pm$ are mass fractions defined in Remark \ref{rem:massf}.

Now, the projection matrix $\P$ can be immediately computed by (\ref{eq:pcomput}):
\begin{equation*}
  \P = \begin{pmatrix}
  		  1 & 0 & 0 & 0 & 0 & 0 \\
  		  0 & 1 & 0 & 0 & 0 & 0 \\
  		  0 & 0 & m^+\Id & m^-\Id & 0 & 0 \\
  		  0 & 0 & 0 & 0 & 1 & 0 \\
  		  0 & 0 & 0 & 0 & 0 & 1 \\
  		 \end{pmatrix}.
\end{equation*}
Finally, after computing all matrix products $\P\A^{-1}(U)\B(U)\J$, $\P\A^{-1}(U)\div\T(U)$, $\P\A^{-1}(U)\S(U)$ present in Equation (\ref{eq:Peq2}), we obtain the following model:
\begin{eqnarray}\label{eq:eq1}
  \alpha^\pm\rho^\pm T^\pm\od{s^\pm}{t} &=& \alpha^\pm\ttau^\pm:\grad\u^\pm - \div(\alpha^\pm\q^\pm), \\
  \rho\od{\u}{t} + \grad p &=& \rho\g + \div\ttau,
\end{eqnarray}
\begin{multline}
  \Pi\bigl(\od{p}{t} + \rho c_s^2\div\u\bigr) = \frac{\rho^-(c_s^-)^2}{\theta^+\rho^+}\bigl(\alpha^+\ttau^+:\grad\u - \div(\alpha^+\q^+)\bigr) \\ + \frac{\rho^+(c_s^+)^2}{\theta^-\rho^-}\bigl(\alpha^-\ttau^-:\grad\u - \div(\alpha^-\q^-)\bigr),
\end{multline}
\begin{multline}
  \Pi\bigl(\od{\alpha^+}{t} + \alpha^+\alpha^-\delta\div\u\bigr) =  \frac{\alpha^-}{\theta^+\rho^+}\bigl(\alpha^+\ttau^+:\grad\u - \div(\alpha^+\q^+)\bigr) \\ - \frac{\alpha^+}{\theta^-\rho^-}\bigl(\alpha^-\ttau^-:\grad\u - \div(\alpha^-\q^-)\bigr), \label{eq:eq3}
\end{multline}
where $\rho = \alpha^+\rho^+ + \alpha^-\rho^-$ is the mixture density (\ref{eq:mixtdens}), $c_s^2$ physically is the sound velocity in the mixture:
\begin{equation}\label{eq:rhocs2}
  \Pi\rho c_s^2 := \rho^+(c_s^+)^2\rho^-(c_s^-)^2.
\end{equation}
To shorten the notation, we also introduced a quantity $\delta$ defined by
\begin{equation}\label{eq:delta}
  \Pi\delta := \rho^+(c_s^+)^2 - \rho^-(c_s^-)^2
\end{equation}
and $\Pi := \alpha^-\rho^+(c_s^+)^2 + \alpha^+\rho^-(c_s^-)^2$.

The viscous stress tensor of the mixture is defined as $\ttau := \alpha^+\ttau^+ + \alpha^-\ttau^-$. If both fluids are assumed to be Newtonian (\ref{eq:viscous}), we can express it in closed form:
\begin{equation*}
  \ttau := \lambda\tr\D(\u)\Id + 2\mu\D(\u), \quad
  \lambda := \alpha^+\lambda^+ + \alpha^-\lambda^-, \quad
  \mu := \alpha^+\mu^+ + \alpha^-\mu^-.
\end{equation*}
Similarly, the thermal flux $\q$ in the mixture is given by this formula:
\begin{equation*}
  \q := \alpha^+\q^+ + \alpha^-\q^-.
\end{equation*}

For practical computations it is better to rewrite Equations (\ref{eq:eq1}) -- (\ref{eq:eq3}) in conservative form \cite{Lax1973,Godlewski1996} which is valid for discontinuous solutions as well. After repeating in inverse sense computations from Theorem (\ref{thm:noncons}) one obtains the following system:
\begin{eqnarray}\label{eq:four1}
  \dt(\alpha^\pm\rho^\pm) + \div(\alpha^\pm\rho^\pm\u) = 0, \\
  \dt(\rho\u) + \div(\rho\u\otimes\u + p\Id) = \rho\g + \div\ttau, \label{eq:four2} \\
  \dt(\rho E) + \div(\rho H\u) = \rho\u\cdot\g + \div(\ttau\u) - \div\q. \label{eq:four3}
\end{eqnarray}

If we neglect viscous stress $\ttau$ and thermal flux $\q$, we will obtain the so-called four equations model recently proposed by F.~Dias, D.~Dutykh and J.-M.~Ghidaglia \cite{Dutykh2007a,Dias2008a,Dias2008,DDG2008} for simulation of violent aerated flows. Formal computations presented in this study can be considered as a step towards justification of the four equations model. It can be also shown \cite{Dias2008}, that the advection operator of Equations (\ref{eq:four1}) -- (\ref{eq:four3}) is hyperbolic for quite general equations of state (\ref{eq:EOS}). Hence, derived here four equations model is very attractive from modeling and computational point of view.


\section{Invariant regions}\label{sec:inv}

Consider only the hyperbolic part of the system (\ref{eq:four1}) -- (\ref{eq:four3}) together with source terms due to gravity:
\begin{eqnarray}
  \dt(\alpha^\pm\rho^\pm) + \div(\alpha^\pm\rho^\pm\u) &=& 0, \label{eq:fourHyp1} \\
  \dt(\rho\u) + \div(\rho\u\otimes\u + p\Id) &=& \rho\g, \label{eq:fourHyp2} \\
  \dt(\rho E) + \div(\rho H\u) &=& \rho\u\cdot\g. \label{eq:fourHyp3}
\end{eqnarray}

In this section we would like to study the preservation of invariant regions under the dynamics of the system (\ref{eq:fourHyp1}) -- (\ref{eq:fourHyp3}). This property is very important in practice for stability of computations. Namely, from physical sense and from definition (\ref{eq:alphaDef}) of volume fractions $\alpha^\pm$ it follows that $0\leq\alpha^\pm\leq 1$, $\forall (\x,t)\in \Omega\times [0, T]$. The question we address here is whether the region $0\leq\alpha^\pm\leq 1$ remains invariant under the system (\ref{eq:fourHyp1}) -- (\ref{eq:fourHyp3}) dynamics. This property can be checked only for $\alpha^+$, for example, since $\alpha^+ + \alpha^- = 1$. If the answer is negative, the model can be hardly applied to any practical situation. Positive result was already proved for a barotropic model with two velocities in \cite{Bresch2008}. In the same work of Bresch et al. (2009) it was also shown that two-phase mixtures of compressible/incompressible fluids do not preserve the invariant regions.

The notion of invariant regions is due to Chueh, Conley and Smoller \cite{Chueh1977}. Independently it was also introduced by Weinberger \cite{Weinberger1975} for scalar parabolic equations.

In order to handle this problem, it is more convenient to rewrite Equations (\ref{eq:fourHyp1}) -- (\ref{eq:fourHyp3}) in terms of the so-called physical variables $W = ^{t}(\u, p, \alpha^+, s)$. The derivation of the equation for the velocity $\u$ evolution is straightforward. Equations for thermodynamical variables $p$, $\alpha^+$ and $s$ are trickier to obtain. However, it was already done in \cite{DDG2008,Dias2008} and partially above (see Equations (\ref{eq:eq1}) -- (\ref{eq:eq3})):
\begin{lem}
Smooth solutions to Equations (\ref{eq:fourHyp1}) -- (\ref{eq:fourHyp3}) satisfy the following quasilinear system:
\begin{eqnarray}
  \dt\u + \u\cdot\grad\u + \frac{\grad p}{\rho} &=& \g, \label{eq:eqphys1} \\
  \dt p + \u\cdot\grad p + \rho c_s^2\div\u &=& 0, \\
  \dt\alpha^+ + \u\cdot\grad\alpha^+ + \alpha^+\alpha^-\delta\div\u &=& 0, \\
  \dt s + \u\cdot\grad s &=& 0, \label{eq:eqphys3}
\end{eqnarray}
where $\rho c_s^2$ and $\delta$ are defined in (\ref{eq:rhocs2}) and (\ref{eq:delta}) respectively.
\end{lem}

\begin{pf}
For the proof see, for example, Dias et al. (2009) \cite{Dias2008} where the authors worked with an auxiliary variable $\alpha := \alpha^+ - \alpha^-$. Taking into account the relation $\alpha^+ + \alpha^- = 1$, one can express the volume fractions in terms of $\alpha$: $\alpha^\pm = \displaystyle \frac{1\pm\alpha}{2}$. Consequently, $\alpha^+\alpha^-=\displaystyle \frac{1-\alpha^2}{4}$.
\end{pf}

\begin{cor}
System (\ref{eq:eqphys1}) -- (\ref{eq:eqphys3}) which governs the evolution of physical variables can be recast under the following matricial form:
\begin{equation}\label{eq:sysM}
	\pd{W}{t} + \M(W)\pd{W}{x} = \S (W),
\end{equation}
where $W = (\u, p, \alpha^+, s)$, source terms $\S(W)={}^t(\g, 0, 0, 0)$ and if $\n = (n_1, n_2, n_3)$ is a normal direction, $u_n = \u\cdot\n$ is a normal velocity, then
\begin{equation*}
		\M(W)\n = \sum_{i=1}^3 M_i(W)n_i = 
					\begin{pmatrix}
						u_n & 0 & 0 & \frac{1}{\rho}n_1 & 0 & 0 \\
						0 & u_n & 0 & \frac{1}{\rho}n_2 & 0 & 0 \\
						0 & 0 & u_n & \frac{1}{\rho}n_3 & 0 & 0 \\
						\rho c_s^2 n_1 & \rho c_s^2 n_2 & \rho c_s^2 n_3 & u_n & 0 & 0 \\
						\alpha^+\alpha^-\delta n_1 & \alpha^+\alpha^-\delta n_2 & \alpha^+\alpha^-\delta n_3 & 0 & u_n & 0 \\
						0 & 0 & 0 & 0 & 0 & u_n \\
					 \end{pmatrix}.
\end{equation*}
\end{cor}

Now we can state the main result:
\begin{thm}
For smooth solutions to Equations (\ref{eq:fourHyp1}) -- (\ref{eq:fourHyp3}), the region $0 \leq \alpha^\pm \leq 1$ remains invariant under the system dynamics.
\end{thm}

\begin{pf}
The proof is based on the theory of J. Smoller \cite{Smoller1994}. We know that the value $\alpha^-=0$ corresponds to $W_5 = \alpha^+ = 1$. Obviously, the case $\alpha^+ = 0$ can be treated similarly with the same conclusions. The matrices $M_i(W)$ are smooth functions of $W$ in the neighborhood of $W$ which satisfies $W_5 \in [0,1]$. We would like to check whether the boundary $W_5=1$ is invariant under the dynamics of (\ref{eq:sysM}). According to the theory of invariant regions by Chueh, Conway and Smoller (\cite{Chueh1977}), this will be the case if and only if $d(W_5-1) = dW_5 = (0,0,0,0,1,0)$ is a left eigenvector of matrices $M_i(\u, p , 1, s)$ for all admissible values of $\u$, $p$ and $s$. By straightforward computation one can easily check that 
\begin{equation*}
(0,0,0,0,1,0)\cdot M_i(\u, p , 1, s) = \vec{u}_i (0,0,0,0,1,0).
\end{equation*}
The theorem is shown now.
\end{pf}

\begin{rem}
The present result shows also that if negative values of volume fractions $\alpha^\pm$ are reported in simulations, they are due to some numerical instabilities and have nothing to do with mathematical properties of the four equations model (\ref{eq:fourHyp1}) -- (\ref{eq:fourHyp3}).
\end{rem}


\section{Incompressible limit}\label{sec:Mach}

In practice, there are many situations when governing equations can be further simplified by filtering out acoustic effects. For various combustion models it was done by A. Majda and his collaborators \cite{Majda1982,Klainerman1982,Majda1985}. More recently these techniques were applied to some two-phase models \cite{Kumbaro2000,Dellacherie2005}, \cite{Dellacherie2007}.

In the present section we derive the incompressible limit of the four equations model (\ref{eq:four1}) -- (\ref{eq:four3}). For the sake of simplicity we will neglect thermal fluxes which are not important for the exposition below and do not directly affect acoustic waves propagation. Hence, in this section we consider the following set of equations in nonconservative form for convenience:
\begin{eqnarray}\label{eq:mach1}
  \dt(\alpha^\pm\rho^\pm) + \div(\alpha^\pm\rho^\pm\u) = 0, \\
  \rho\dt\u + \rho(\u\cdot\grad)\u + \grad p = \rho\g + \div\ttau, \\
  \rho\dt e + \rho\u\cdot\grad e + p\div\u = \ttau:\grad\u, \label{eq:mach3}
\end{eqnarray}
where we expressed the energy balance in terms of the internal energy $e$.

In order to estimate the relative importance of various terms, we introduce dimensionless variables. In this section we use the special low Mach number scaling which reveals acoustic effects magnitude. The characteristic length, time, and velocity scales are denoted by $\ell$, $t_0$ and $U_0$ respectively. For example, $\ell$ can be chosen as a linear dimension of the fluid domain $\Omega$. The density, viscosity and sound velocity scales are chosen to be those of the heavy fluid, i.e. $\rho_0^+$, $\nu_0^+$ and $c_{0s}^+$ correspondingly. Since we are interested in acoustic effects, the natural pressure scale is given by $\rho_0^+(c_{0s}^+)^2$. If we summarize these remarks, dependent and independent dimensionless variables (denoted with primes) are defined as:
\begin{equation*}
  \x' := \frac{\x}{\ell}, \quad t' := \frac{t}{t_0}, \quad  
  (\rho^\pm)' := \frac{\rho^\pm}{\rho_0^+}, \quad 
  (\mu^\pm)' := \frac{\mu^\pm}{\rho_0^+\nu_0^+}, 
\end{equation*}
\begin{equation*}
	\u' := \frac{\u}{U_0}, \quad p' := \frac{p}{\rho_0^+(c_{0s}^+)^2}, \quad
	e' := \frac{e}{U_0^2}.
\end{equation*}
Since the volume fraction is dimensionless by definition (\ref{eq:alphaDef}) we keep this variable unchanged.

After dropping the tildes, nondimensional system (\ref{eq:mach1}) -- (\ref{eq:mach3}) of equation becomes:
\begin{eqnarray}\label{eq:mach1ad}
  \St\dt(\alpha^\pm\rho^\pm) + \div(\alpha^\pm\rho^\pm\u) = 0, \\
  \St\rho\dt\u + \rho(\u\cdot\nabla)\u + \frac{1}{\Ma^2}\nabla p = \frac{1}{\Fr^2}\rho\g + \frac{1}{\Re}\div\ttau, \label{eq:mach2ad} \\
  \St\rho\dt e + \rho\u\cdot\grad e + \frac{1}{\Ma^2}p\div\u = \frac{1}{\Re}\ttau:\grad\u, \label{eq:mach3ad}
\end{eqnarray}
where several scaling parameters appear:
\begin{itemize}
  \item \textbf{Strouhal number} $\St := \displaystyle\frac{\ell}{U_0 t_0}$. In this study we will assume the Strouhal number to be equal to one $\St \equiv 1$, i.e. $t_0 = \displaystyle\frac{\ell}{U_0}$.
  \item \textbf{Mach number} $\Ma := \displaystyle\frac{U_0}{c_{0s}^+}$ which measures the relative importance of the flow speed and the sound speed in the medium.
  \item \textbf{Froude number} $\Fr := \displaystyle\frac{U_0}{\sqrt{g\ell}}$ compares inertia and gravity forces.
  \item \textbf{Reynolds number} $\Re := \displaystyle\frac{U_0\ell}{\nu_0^+}$ gives the measure of the ratio of inertial to viscous forces.
\end{itemize}
In this section we will consider the asymptotic limit as $\Ma\to 0$. Consequently, all physical variables $\alpha^\pm$, $\rho^\pm$, $p$, $\u$ and $e$ are expanded in formal series in powers of the Mach number:
\begin{equation}\label{eq:expansion}
  \phi = \phi_0 + \Ma\:\phi_1 + \Ma^2\phi_2 + \ldots, \qquad
  \phi \in \{\alpha^\pm, \rho^\pm, p, \u, e \}.
\end{equation}
Formal expansion (\ref{eq:expansion}) is then substituted into the system (\ref{eq:mach1ad}) -- (\ref{eq:mach3ad}). Our goal is to derive a system which governs the evolution of physical variables at the lowest order in the Mach number (e.g. $\u_0$, $\alpha_0^\pm$, etc.). Equation (\ref{eq:mach2ad}) at orders $\Ma^{-2}$ and $\Ma^{-1}$ leads:
\begin{equation*}
  \nabla p_0 = \nabla p_1 = 0.
\end{equation*}
In other words, $p_0 = p_0 (t)$ and $p_1 = p_1(t)$ are only functions of time. Internal energy balance equation (\ref{eq:mach3ad}) gives us the incompressibility constraint $\div\u_0 = 0$ at the leading order $\Ma^{-2}$ and, correspondingly, $\div\u_1 = 0$ at the order $\Ma^{-1}$. 

Relation (\ref{eq:alpha}) after asymptotic expansion (\ref{eq:expansion}) becomes
\begin{equation*}
  \alpha^+_k (\x,t) + \alpha^-_k (\x,t) = \delta_{0k},
\end{equation*}
where $\delta_{0k}$ is the Kronecker delta symbol equal to $1$ if $k = 0$ and $0$ otherwise. Thus, at the leading order in Mach number we keep usual relations:
\begin{equation*}
  \alpha_0^+ + \alpha_0^- = 1, \qquad
  \rho_0 = \alpha_0^+\rho_0^+ + \alpha_0^-\rho_0^-.
\end{equation*}
Equations (\ref{eq:mach1ad}), (\ref{eq:mach2ad}) at the order $\Ma^0$ become:
\begin{eqnarray}\label{eq:mass0}
  \dt(\alpha^\pm_0\rho^\pm_0) + \div(\alpha^\pm_0\rho^\pm_0\u_0) = 0, \\
  \rho_0\dt\u_0 + \rho_0(\u_0\cdot\nabla)\u_0 + \nabla\pi = \frac{1}{\Fr^2}\rho_0\g + \frac{1}{\Re}\div\ttau_0,
\end{eqnarray}
where by $\pi$ we denote the pressure oscillations $p_2$ at the order $\Ma^2$.

Finally, some information on the behaviour of $\rho_0^\pm$ can be deduced from the Gibbs relation (\ref{eq:Gibbs}) which takes the following dimensionless form:
\begin{equation*}
  T^\pm ds^\pm = de^\pm - \frac{p}{\Ma^2(\rho^\pm)^2}d\rho^\pm.
\end{equation*}
After dividing by $dt$ and expanding the Gibbs relation as in (\ref{eq:expansion}) we get these results:
\begin{equation}\label{eq:drhodt}
  \od{\rho^\pm_0}{t} = 0, \qquad \od{\rho^\pm_1}{t} = 0.
\end{equation}
In particular, it means that if in each phase the density $\rho_{0,1}^\pm$ was initially constant\footnote{This assumption is often made in applications. For example, pure water and pure air are assumed to have constant densities under normal conditions. However, the mixture density can undergo strong variations (e.g. after the wave breaking).}, it remains so under the system dynamics. Moreover, taking into account the first relation in (\ref{eq:drhodt}) and the flow incompressibility constraint $\div\u_0 = 0$, one can easily deduce from the mass conservation Equations (\ref{eq:mass0}) that volume fractions $\alpha_0^\pm$ are also simply transported by the flow:
\begin{equation*}
  \dt\alpha_0^\pm + \u_0\cdot\grad\alpha_0^\pm = 0.
\end{equation*}

If we summarize all the developments made above and turn back to physical variables, we will get the following system of equations which governs an incompressible two-phase flow:
\begin{eqnarray}
  \dt\alpha^\pm + \u\cdot\grad\alpha^\pm = 0, \label{eq:inc1} \\
  \div\u = 0, \label{eq:inc2} \\
  \rho\dt\u + \rho(\u\cdot\nabla)\u + \nabla\pi = \rho\g + \div\ttau, \label{eq:inc3}
\end{eqnarray}
where we dropped the index $0$ to simplify the notation. The mixture density $\rho$ and the viscous stress tensor $\ttau$ are defined as above. This system should be completed by appropriate boundary and initial conditions.

\begin{rem}
Two-fluid incompressible models, such as the system just derived above (\ref{eq:inc1}) -- (\ref{eq:inc3}), are often used in many practical situations such as wave breaking \cite{Chen1999} and others \cite{Scardovelli1999}. For powder-snow avalanches applications, the volume fraction Equation (\ref{eq:inc1}) is completed by a diffusive term according to the Fick's law:
\begin{equation*}
	\dt\alpha^\pm + \u\cdot\grad\alpha^\pm = \div(\nu_f\grad\alpha^\pm).
\end{equation*}
This extra term allows to take into account the mixing between two fluids due to turbulence effects, for example. For more details on this extension we refer to \cite{Joseph1993,Etienne2004,Dutykh2009}.
\end{rem}


\section{Numerical results}\label{sec:num}

In the present section we will consider only the advective part of the system (\ref{eq:four1}) -- (\ref{eq:four3}), i.e. we neglect viscous and thermal fluxes:
\begin{eqnarray}\label{eq:consfour1}
  \dt(\alpha^\pm\rho^\pm) + \div(\alpha^\pm\rho^\pm\u) = 0, \\
  \dt(\rho\u) + \div(\rho\u\otimes\u + p\Id) = \rho\g, \\
  \dt(\rho E) + \div(\rho H\u) = \rho\u\cdot\g. \label{eq:consfour3}
\end{eqnarray}
Actually, the system we consider here is stiffer than the original Equations (\ref{eq:four1}) -- (\ref{eq:four3}) since it does not contain any diffusive effects. The resulting equations have been shown to be hyperbolic for any reasonable equation of state \cite{Dias2008,DDG2008}. For illustrative purposes we assume here that both fluids are governed by stiffened gas type laws:
\begin{equation*}
  p^\pm + \pi^\pm = (\gamma^\pm - 1)\rho^\pm e^\pm, \quad
  e^\pm = c_v^\pm T^\pm + \frac{\pi^\pm}{\gamma^\pm\rho^\pm},
\end{equation*}
where $\gamma^\pm$, $\pi^\pm$ and $c_v^\pm$ are some constants determined by physical properties of pure fluids under consideration. Also we assume that two fluids are in the thermodynamic equilibrium:
\begin{equation*}
 p \equiv p^+ = p^-, \quad T \equiv T^+ = T^-.
\end{equation*}
We refer to \cite{Dutykh2007a,Dias2008,DDG2008} for more details on the construction of this equation of state and the discussion of some properties of such two-phase mixtures.

\subsection{Numerical schemes}

For the numerical study we choose the finite volumes method \cite{Kroner1997,Barth2004} since it is the method of choice for the systems of conservation laws due to its excellent local conservative properties. More precisely, we use the cell-centered approach \cite{Barth1989,Barth1994} which is more natural in our opinion. For simplicity we assume that the system of equations is solved in $\R^2$. However the extension to 3D for cartesian meshes is straightforward. We briefly describe below the discretization procedure adopted in this study. References are also provided if more details of the discretization method are needed.

\subsubsection{Space discretization}

System (\ref{eq:consfour1}) -- (\ref{eq:consfour3}) can be written as
\begin{equation}\label{eq:conslaw}
  \pd{\w}{t} + \div\F(\w) = \S(\w)\,,
\end{equation}
where
\begin{equation*}
  \w = (w_i)_{i=1}^{5} := (\alpha^+\rho^+,\;\; \alpha^-\rho^-,\;\;\rho u_1,\;\;\rho u_2,\;\;\rho E)\,,
\end{equation*}
and, for every $\n = (n_1, n_2) \in\R^2$,
\begin{multline}\label{flux1}
 \F(\w)\cdot\n = (\alpha^+\rho^+\u\cdot \n,\;\; \alpha^-\rho^-\u\cdot \n,\;\; 
 \rho\u\cdot  \n u_1+ p n_1, \\
 \rho \u\cdot \n u_2 + pn_2,\;\; \rho H \u\cdot \n)\,,
\end{multline}
\begin{equation*}
	\S(\w)=(0,\;\; 0,\;\; \rho g_1,\;\; \rho g_2,\;\; \rho\g\cdot\u)\,.
\end{equation*}
Then, the Jacobian matrix $\A(\w)\cdot\n$ is defined by
\begin{equation}\label{mat_jacob}
	\A_n(\w) := \A(\w)\cdot\n=\pd{(\F(\w) \cdot\n)}{\w}\,.
\end{equation}

In order to compute $\A(\w)\cdot\n$, one writes Equation (\ref{flux1}) for $\F(\w)\cdot\n$ in terms of $\w$ and $p$:
\begin{multline*}
	  \F(\w)\cdot\n =  \Bigl(w_1\frac{w_3n_1 + w_4n_2}{w_1+w_2},\;\; w_2\frac{w_3n_1 + w_4n_2}{w_1+w_2},\;\; w_3\frac{w_3n_1 + w_4n_2}{w_1+w_2} + p n_1, \\ w_4\frac{w_3n_1 + w_4n_2}{w_1+w_2} + pn_2,\;\; (w_5 + p)\frac{w_3n_1 + w_4n_2}{w_1+w_2}\Bigr)\,.
\end{multline*}
The Jacobian matrix (\ref{mat_jacob}) then has the following expression:
\begin{eqnarray*}
 \A(\w)\cdot\n & = & \\
 & & \hspace{-3.5cm} \begin{pmatrix}
   u_n\frac{\alpha^-\rho^-}{\rho} & -u_n\frac{\alpha^+\rho^+}{\rho} & \frac{\alpha^+\rho^+}{\rho} n_1 & \frac{\alpha^+\rho^+}{\rho} n_2 & 0 \\
   -u_n\frac{\alpha^-\rho^-}{\rho} & u_n\frac{\alpha^+\rho^+}{\rho} & \frac{\alpha^-\rho^-}{\rho} n_1 & \frac{\alpha^-\rho^-}{\rho} n_2 & 0 \\
   -u_1 u_n+\pd{p}{w_1}n_1 & -u_1 u_n+\pd{p}{w_2}n_1 & u_n + u_1 n_1 + \pd{p}{w_3}n_1 & %
   u_1 n_2 + \pd{p}{w_4} n_1 & \pd{p}{w_5} n_1 \\
   -u_2 u_n+\pd{p}{w_1}n_2 & -u_2 u_n+\pd{p}{w_2}n_2 & u_2 n_1 + \pd{p}{w_3}n_2 & %
   u_n + u_2 n_2 + \pd{p}{w_4} n_2 & \pd{p}{w_5} n_2 \\
   u_n\bigl(\pd{p}{w_1} - H\bigr) & u_n\bigl(\pd{p}{w_2} - H\bigr) & %
   u_n\pd{p}{w_3} + H n_1 & u_n\pd{p}{w_4} + H n_2 & u_n\bigl(1 + \pd{p}{w_5}\bigr) \\
 \end{pmatrix}\,,
\end{eqnarray*}
where $u_n = \u \cdot \n.$

The computational domain $\Omega\subset\R^2$ is decomposed into a set of control volumes $\Tt$ such that $\Omega=\cup_{K\in\Tt} K$. We integrate Equation (\ref{eq:conslaw}) on $K$:
\begin{equation}\label{eq:conservlaw}
	\od{}{t}\int_{K} \w \;d\Omega + \sum_{L\in\N(K)}\int_{K\cap L}\F(\w)\cdot\n_{KL} \;d\sigma = \int_{K} \S(\w) \;d\Omega\,,
\end{equation}
where $\n_{KL}$ denotes the unit normal vector on $K\cap L$ pointing into $L$ and $\N(K) = \set{L\in\Tt: \area(K\cap L) \neq 0}\,.$ Then, setting
\begin{equation*}
 \w_K(t) := \frac{1}{\vol(K)}\int_{K} \w(\x,t) \;d\Omega \;,
\end{equation*}
we approximate (\ref{eq:conservlaw}) by
\begin{equation*}
	\od{\w_K}{t} + \sum_{L\in\N(K)} \frac{\area(L\cap K)}{\vol(K)} \Phi(\w_K, \w_L; \n_{KL}) =  \S_K\;,
\end{equation*}
where the numerical flux 
\begin{equation*}
\Phi(\w_K, \w_L; \n_{KL})\approx\frac{1}{\area(L\cap K)}\int_{K\cap L}
\F(\w)\cdot\n_{KL} \;d\sigma
\end{equation*}
is explicitly computed by the FVCF formula of Ghidaglia {\it et al.} \cite{Ghidaglia2001}:
\begin{multline}\label{eq:FVCF}
\Phi(\w_K, \w_L; \n)=\frac{\F(\w_K) \cdot\n+\F(\w_L) \cdot\n}{2} \\ -\sign(\A_n(\mu(\w_K,\w_L)))\frac{\F(\w_K)\cdot\n - \F(\w_L)\cdot\n}{2}\,.
\end{multline}
Here the Jacobian matrix $\A_n(\mu)$ is defined in (\ref{mat_jacob}), $\mu(\w_K,\w_L)$ is an arbitrary mean between $\w_K$ and $\w_L$ and $\sign(M)$ is the matrix whose eigenvectors are those of $M$ but whose eigenvalues are the signs of that of $M$. In this section we did not deal with boundary conditions. We refer to \cite{Ghidaglia2005} for more details.

\subsubsection{Higher order extension}

In the previous section we described the first order scheme which might be too diffusive for most practical applications. That is why we present here a higher-order extension which is a variant of MUSCL\footnote{Acronym MUSCL stands for Monotone Upstream-centered Scheme for Conservation Laws \cite{Leer1979}} limiting technique \cite{Kolgan1975,Kolgan1972,Leer1979,Leer2006}. This numerical method ensures stability and non-oscillatory behaviour of numerical solutions. To describe this scheme we switch to Cartesian notation since a bigger stencil is needed for the gradient reconstruction procedure. In this notation $\w_i$ denotes the average of conservative variables in the cell $i$ and $\w_{i+\frac12}^{L,R}$ denotes respectively reconstructed left and right states at cell faces. According to the method adopted in the current study \cite{Barth1994,Barth2004,Thornber2007}, cell faces values are computed as:
\begin{equation}\label{eq:reconstruct1}
 \w_{i+\frac12}^L = \w_i + \frac14\Bigl((1-\kappa)\phi(r_i^L)(\w_i - \w_{i-1})
 + (1+\kappa)\phi\bigl(\frac{1}{r_i^L}\bigr)(\w_{i+1}-\w_i)\Bigr),
\end{equation}
\begin{equation}\label{eq:reconstruct2}
 \w_{i+\frac12}^R = \w_{i+1} - \frac14\Bigl((1-\kappa)\phi(r_i^R)(\w_{i+2} - \w_{i+1}) + (1+\kappa)\phi\bigl(\frac{1}{r_i^R}\bigr)(\w_{i+1}-\w_i)\Bigr),
\end{equation}
where $\kappa\in [-1, 1)$ is a free parameter and
\begin{equation*}
  r_i^L := \frac{\w_{i+1} - \w_i}{\w_i - \w_{i-1}}, \quad
  r_i^R := \frac{\w_{i+1} - \w_i}{\w_{i+2} - \w_{i-1}}.
\end{equation*}
Then, reconstructed values $\w_{i+\frac12}^{L,R}$ are used to compute the numerical flux (\ref{eq:FVCF}) of the FVCF scheme.

The function $\phi(r)$ is called the limiter function and is incorporated to obtain non-oscillatory resolution of discontinuities and steep gradients \cite{Boris1973,Sweby1984}. We tested in practice two following limiter functions:
\begin{itemize}
  \item M3 limiter proposed in \cite{Zoltak1998,Thornber2007}: 
  \begin{equation}\label{eq:M3}
  	\phi_{M3}(r) = 1 - \Bigl(1 + \frac{2Nr}{1+r^2}\Bigr)
  	\Bigl(1-\frac{2r}{1+r^2}\Bigr)^N, \quad N = 2,
  \end{equation}
  
  \item MinMod limiter with compression parameter \cite{Barth1994,Barth2004}: 
  \begin{equation}\label{eq:minmod}
  	\phi_{MM}(r) = \max\{0, \min\{r, \beta\}\}, \quad 
  	\beta\in \bigl(1,\frac{3-\kappa}{1-\kappa}\bigr].
  \end{equation}
\end{itemize}

In the smooth unlimited form ($\phi(r)\equiv 1$), the truncation error $\varepsilon$ for a scalar conservation law with the smooth flux $f(u)$ is given by \cite{Barth1994,Barth2004}:
\begin{equation*}
  \varepsilon = -\frac{\bigl(\kappa - \frac13\bigr)}{4}\Delta x^2 \pd{^3 f(u)}{x^3}.
\end{equation*}
Consequently, for $\kappa = \frac13$, the scheme provides theoretically in \textit{smooth regions} an overall spatial discretization with $\O (\Delta x^3)$ error. In computations presented below, we use the optimal choice of this parameter together with M3 limiter (\ref{eq:M3}).

\begin{rem}
If the limiter function is symmetric, i.e.
\begin{equation*}
  \frac{\phi(r)}{r} \equiv \phi\bigl(\frac{1}{r}\bigr),
\end{equation*}
the reconstruction formulas (\ref{eq:reconstruct1}), (\ref{eq:reconstruct2}) are independent of the parameter $\kappa$ and degenerate to the classical MUSCL2 scheme. It is straightforward to show that limiter functions (\ref{eq:M3}) and (\ref{eq:minmod}) are not symmetric.
\end{rem}

\subsubsection{Time discretization}

In the previous section we briefly described the spatial discretization procedure with some finite-volume scheme. It is a common practice in solving time-dependent PDEs to first discretize the spatial variables. This approach is called a method of lines:
\begin{equation*}
  \pd{\w}{t} + \div \F(\w) = \S(\w) \stackrel{\mbox{FV}}{\Longrightarrow} \w_t = \L(\w),
\end{equation*}
where $\L(\w)$ is a space discretization operator. In order to obtain a fully discrete scheme, we have to discretize the time evolution operator. In the present work we decided to retain the so-called Strong Stability-Preserving (SSP) time discretization methods described in \cite{Shu1988,Gottlieb2001,Spiteri2002}. Historically these methods were initially called Total Variation Diminishing (TVD) time discretizations. However, this term is not completely correct. In computations presented below we use the following third order four-stage SSP-RK(3,4) scheme with CFL $ = 2$:
\begin{eqnarray*}
    \w^{(1)} = \w^{(n)} + \frac12\Delta t \L(\w^{(n)}), \\
    \w^{(2)} = \w^{(1)} + \frac12\Delta t \L(\w^{(1)}), \\
    \w^{(3)} = \frac23 \w^{(n)} + \frac13 \w^{(2)} + \frac16\Delta t \L(\w^{(n)}), \\
    \w^{(n+1)} = \w^{(3)} + \frac12\Delta t \L(\w^{(3)}),
\end{eqnarray*}
where $\w^{(n)} = \w(\cdot, t_n)$ and $\w^{(n+1)} = \w(\cdot, t_{n+1})$.

The time step is chosen adaptively to satisfy the following stability condition \cite{Courant1967}:
\begin{equation*}
  \Delta t \leq 
  \frac{\mathrm{CFL}\cdot\Delta x_{min}}{\max\{|u+c_s|_{max},|u-c_s|_{max}\}},
\end{equation*}
where $\Delta x_{min}$ is the minimal mesh spacing, $u$ is the velocity and $c_s$ is the sound speed in the mixture (\ref{eq:rhocs2}).

\begin{rem}
Presented above time step restriction is used together with the first order finite volumes scheme. The application of the MUSCL3 scheme requires the following modification of the CFL condition:
\begin{equation*}
  \Delta t \leq 
  \frac{1-\kappa}{2-\kappa}\cdot\frac{\mathrm{CFL}\cdot
  \Delta x_{min}}{\max\{|u+c_s|_{max},|u-c_s|_{max}\}}.
\end{equation*}
\end{rem}

\subsection{Two-fluid Sod shock tube problem}

\begin{figure}
	\centering
	\scalebox{1} 
	{
	\begin{pspicture}(0,-2.2629688)(10.3,2.2629688)
	\definecolor{color0b}{rgb}{0.8,0.8,1.0}
	\psframe[linewidth=0.04,dimen=outer,fillstyle=solid,fillcolor=color0b](10.3,1.5445312)(0.02,-1.3154688)
	\psline[linewidth=0.04cm](5.22,1.5245312)(5.22,-1.2554687)
	\psline[linewidth=0.04cm,arrowsize=0.05291667cm 2.0,arrowlength=1.4,arrowinset=0.4]{<->}(0.02,-1.7154688)(10.22,-1.6954688)
	\usefont{T1}{ptm}{m}{n}
	\rput(5.001406,-2.0854688){$L = 2$}
	\psline[linewidth=0.04cm,arrowsize=0.05291667cm 2.0,arrowlength=1.4,arrowinset=0.4]{<->}(0.0,1.7245313)(5.2,1.7445313)
	\usefont{T1}{ptm}{m}{n}
	\rput(2.7214062,2.0745313){$\ell = 1$}
	\usefont{T1}{ptm}{m}{n}
	\rput(2.5114062,0.31453124){$\alpha^+,\;\; \rho^+,\;\; p^+$}
	\usefont{T1}{ptm}{m}{n}
	\rput(7.4714065,0.29453126){$\alpha^-,\;\; \rho^-,\;\; p^-$}
	\usefont{T1}{ptm}{m}{n}
	\rput(2.6414063,-0.36546874){$u = 0$}
	\usefont{T1}{ptm}{m}{n}
	\rput(7.4214063,-0.34546876){$u = 0$}
	\end{pspicture} 
	}
	\caption{Sketch of the initial condition for the Sod shock tube test case.}
	\label{fig:SodSketch}
\end{figure}
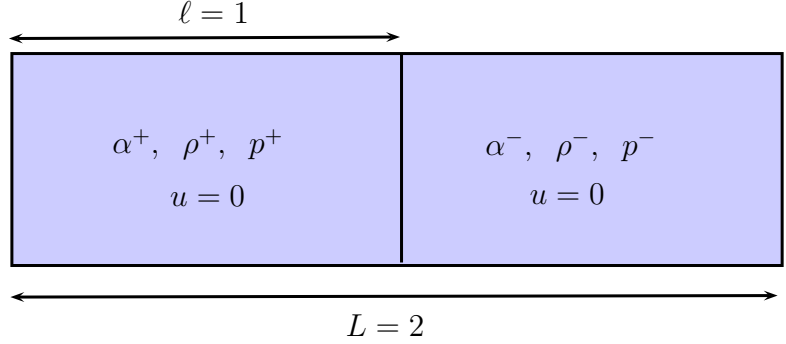

We present here a two-fluid generalization of the classical Sod shock tube problem \cite{Sod1978}. The sketch of the initial condition is given on Figure \ref{fig:SodSketch}. The gravity force is not taken into account in this test case, i.e. $\g = \vec{0}$. Values of all parameters used in this computation are given in Tables \ref{tab:params1} and \ref{tab:EOS1}.

\begin{table}
  \begin{center}
    \begin{tabular}{r||c|c|c|c}
     	 & Velocity, $u$ & Density, $\rho^\pm$ & Volume fraction, $\alpha^+$ & Pressure, $p$ \\
     \hline\hline
     Heavy fluid, $+$ & 0.0 & 1.0 & 0.98 & 1.0 \\
     \hline
     Light fluid, $-$ & 0.0 & 0.125 & 0.02 & 0.1 \\
    \end{tabular}
  \end{center}
  \caption{Initial condition parameters for the two-fluid Sod shock tube problem.}
  \label{tab:params1}
\end{table}

\begin{table}
  \begin{center}
    \begin{tabular}{r||c|c|c}
      & $\gamma^\pm$ & $\pi^\pm$, Pa & $c_v^\pm$, $\frac{J}{kg\cdot K}$ \\
     \hline\hline
     Heavy fluid, $+$ &  2.6 & 0.0 & 661.0 \\
     \hline
     Light fluid, $-$ & 1.4 & 0.0 & 661.0 \\
    \end{tabular}
  \end{center}
  \caption{Equation of state parameters for the two-fluid Sod shock tube problem.}
  \label{tab:EOS1}
\end{table}

\begin{figure}
  \centering
  \subfigure[FVCF 1$^{\mathrm{st}}$ order]%
  {\includegraphics[width=0.49\textwidth]{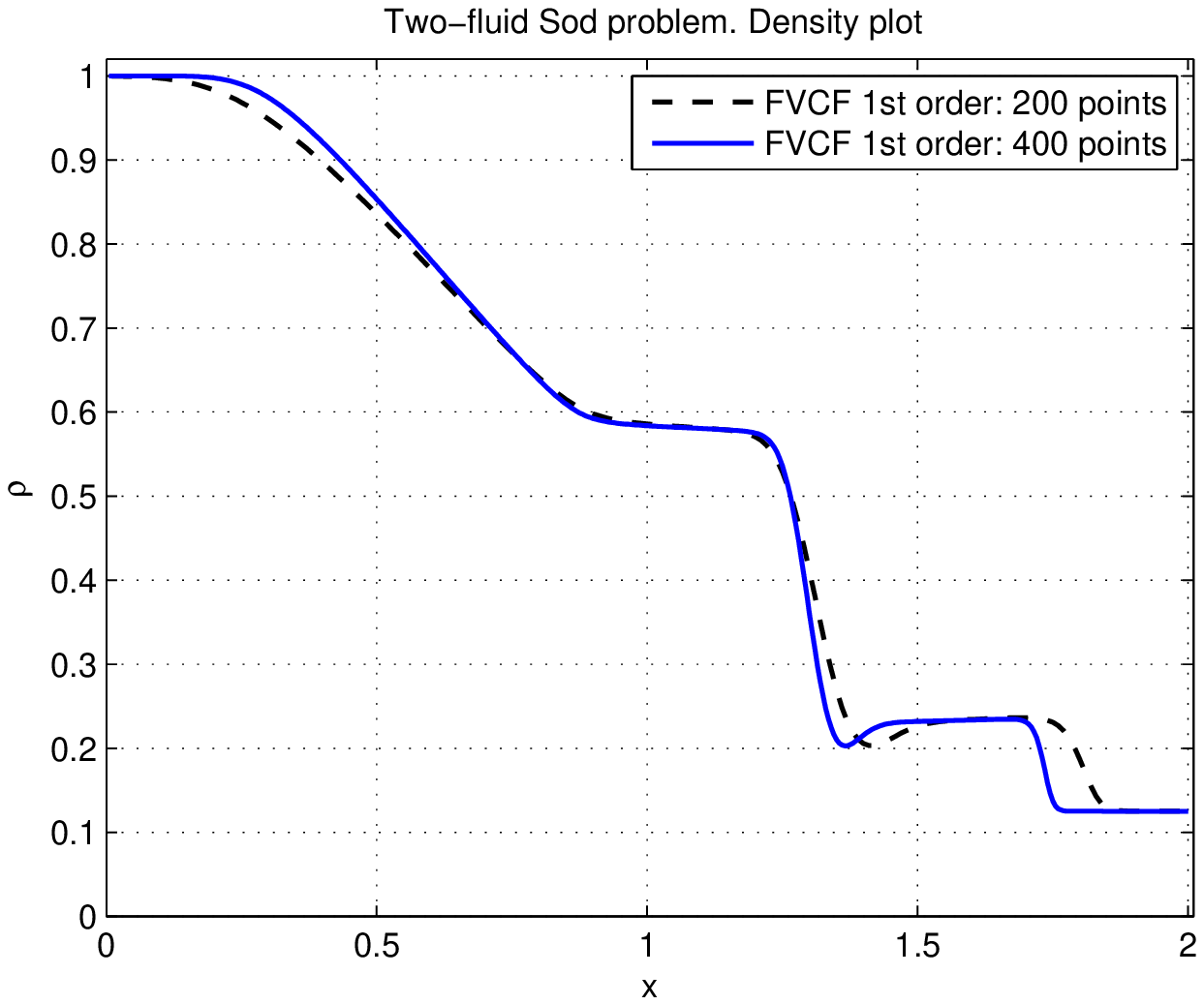}}
  \subfigure[MUSCL 3]%
  {\includegraphics[width=0.49\textwidth]{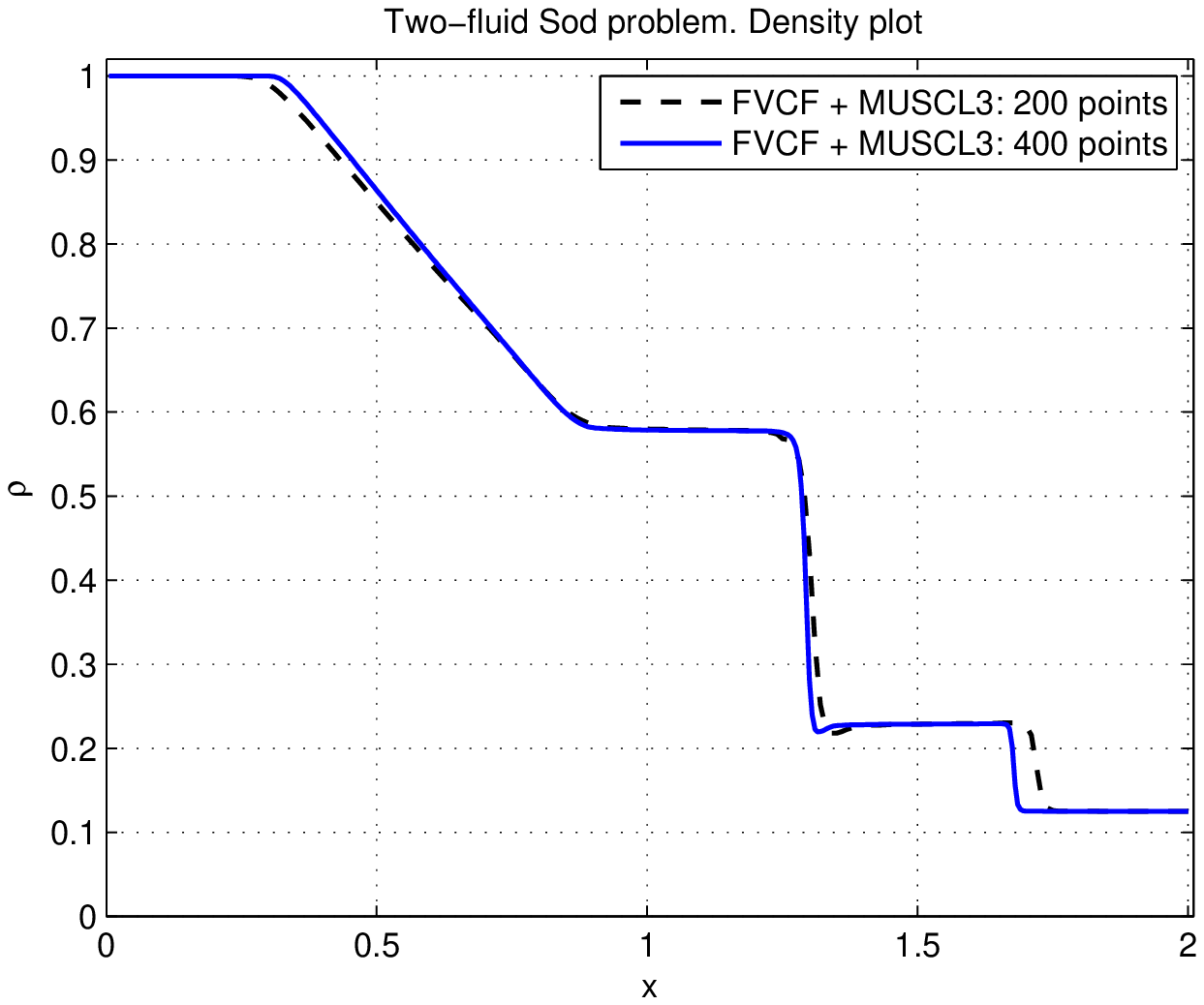}}
  \caption{Convergence of the solution with $h$-refinement}
  \label{fig:hrefine}
\end{figure}

\begin{figure}
  \centering
  \subfigure[$N = 200$]%
  {\includegraphics[width=0.49\textwidth]{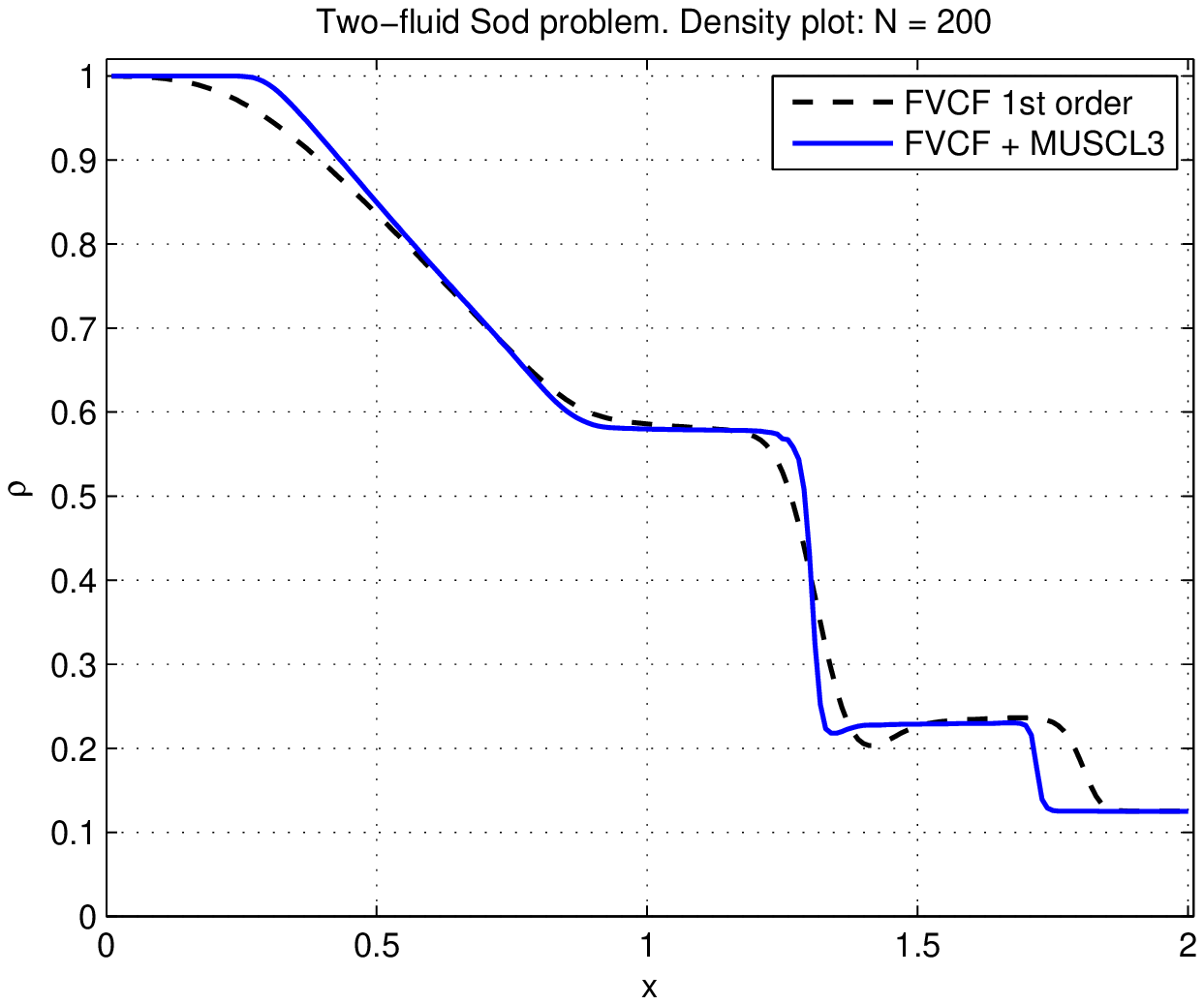}}
  \subfigure[$N = 400$]%
  {\includegraphics[width=0.49\textwidth]{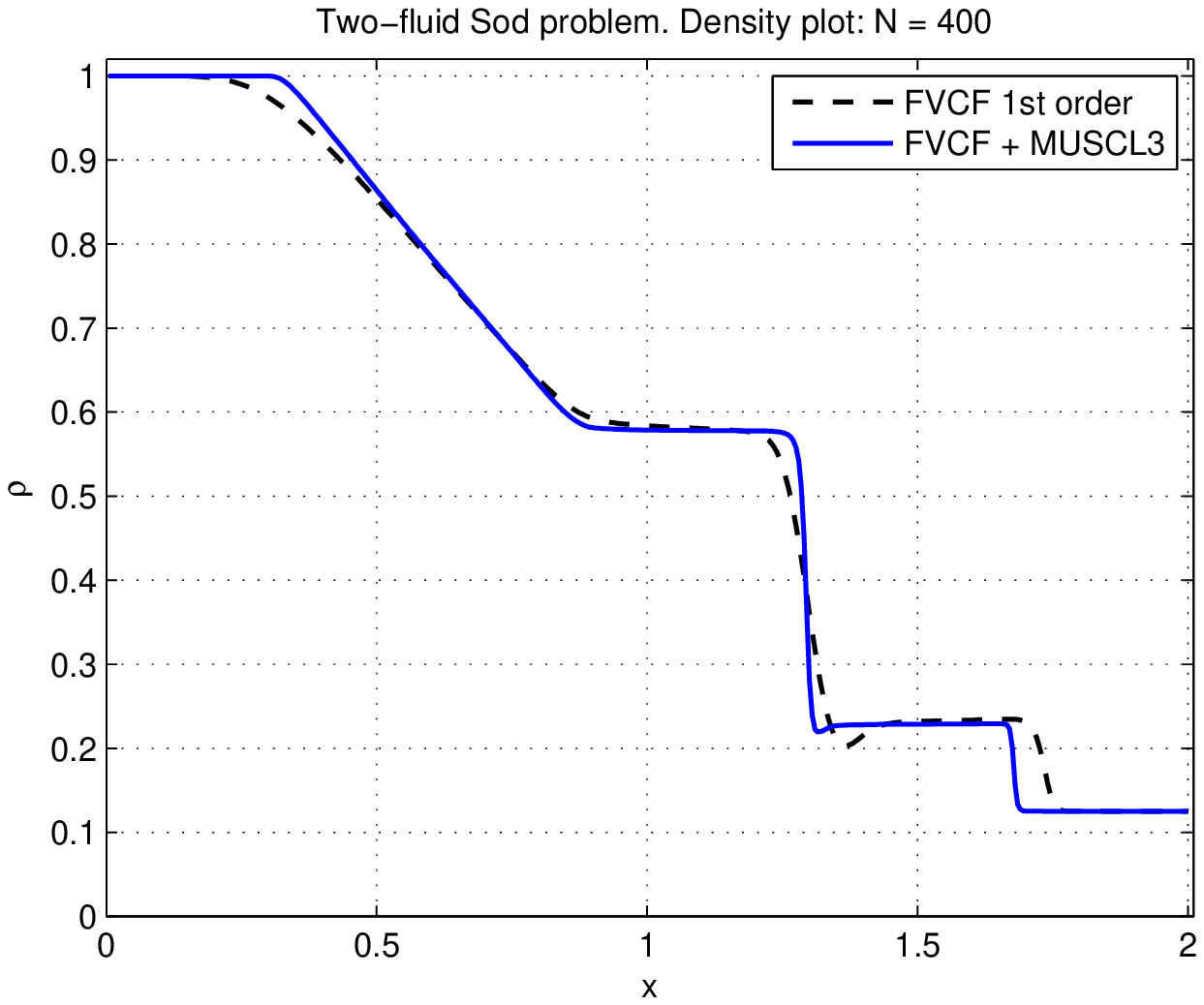}}
  \caption{Convergence of the solution with $p$-refinement}
  \label{fig:prefine}
\end{figure}

The simulation was stopped at time $T = 0.4$ s and the computation results are presented on Figures \ref{fig:hrefine} and \ref{fig:prefine}. On all Figures we show the total density $\rho$ plots and perform a comparison between FVCF and MUSCL3 schemes. Figure \ref{fig:hrefine} shows the behaviour of the solution when the mesh is refined. Left image shows the results of the first order FVCF scheme, while the right one refers to the MUSCL3 extension. The improvement of the solution is obvious in both cases. In particular, we would like to underline the excellent shock resolution by the higher order scheme (Figure \ref{fig:hrefine} (b)).

We perform also the comparison between two schemes (see Figure \ref{fig:prefine}) for the same mesh size. As it was expected, higher order scheme greatly improves the sharp resolution of discontinuities for nearly comparable CPU time. Thus, there is an obvious interest in using MUSCL type schemes for practical applications.

This test case clearly shows the convergence of the numerical solution when the mesh is refined ($h$-convergence), but also with respect to the scheme order ($p$-convergence). Consequently, it validates our numerical code. In all subsequent computations we use the MUSCL3 scheme with the optimal choice of the parameter $\kappa = \frac13$.

\subsection{Water drop test case}

\begin{figure}
	\centering
	\scalebox{1} 
	{
	\begin{pspicture}(0,-3.08375)(5.22,3.04375)
	\definecolor{color1g}{rgb}{0.8,0.8,1.0}
	\definecolor{color1f}{rgb}{0.6,0.6,1.0}
	\definecolor{color51b}{rgb}{0.2,0.4,1.0}
	\psframe[linewidth=0.04,dimen=outer,fillstyle=gradient,gradlines=2000,gradbegin=color1g,gradend=color1f,gradmidpoint=1.0](5.22,3.04375)(0.0,-2.17625)
	\psline[linewidth=0.04cm,arrowsize=0.05291667cm 2.0,arrowlength=1.4,arrowinset=0.4]{<->}(0.0,-2.55625)(5.2,-2.53625)
	\usefont{T1}{ptm}{m}{n}
	\rput(2.4514062,-2.90625){$\ell$}
	\psframe[linewidth=0.04,dimen=outer,fillstyle=solid,fillcolor=color51b](3.56,2.02375)(1.84,0.32375)
	\psline[linewidth=0.04cm,arrowsize=0.05291667cm 2.0,arrowlength=1.4,arrowinset=0.4]{<->}(1.82,0.04375)(3.54,0.04375)
	\psdots[dotsize=0.12](2.7,1.18375)
	\usefont{T1}{ptm}{m}{n}
	\rput(2.75,1.49375){$(x_0, y_0)$}
	\usefont{T1}{ptm}{m}{n}
	\rput(2.8214064,-0.24625){$r_0$}
	\usefont{T1}{ptm}{m}{n}
	\rput(1.8014063,-1.62625){$\alpha^-, \rho^-$}
	\psline[linewidth=0.04cm,arrowsize=0.05291667cm 2.0,arrowlength=1.4,arrowinset=0.4]{->}(5.84,1.88375)(5.84,0.30375)
	\usefont{T1}{ptm}{m}{n}
	\rput(6.2214065,1.05375){$g$}
	\end{pspicture}
	}
	\caption{Sketch of the initial condition and computational domain geometry for the water drop test case. In our simulation we took the following values of geometrical parameters: $\ell = 1.0$ (the domain is square), $r_0 = 0.15$, $x_0 = 0.5$ and $y_0 = 0.7$. The gravity acceleration is taken to be $g = 10$ m/s$^2$.}
	\label{fig:drop}
\end{figure}
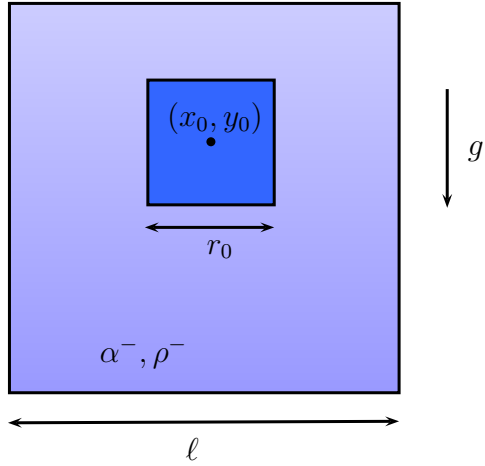

\begin{figure}
  \centering
  \subfigure[$t = 0.1$ s]%
  {\includegraphics[width=0.49\textwidth]{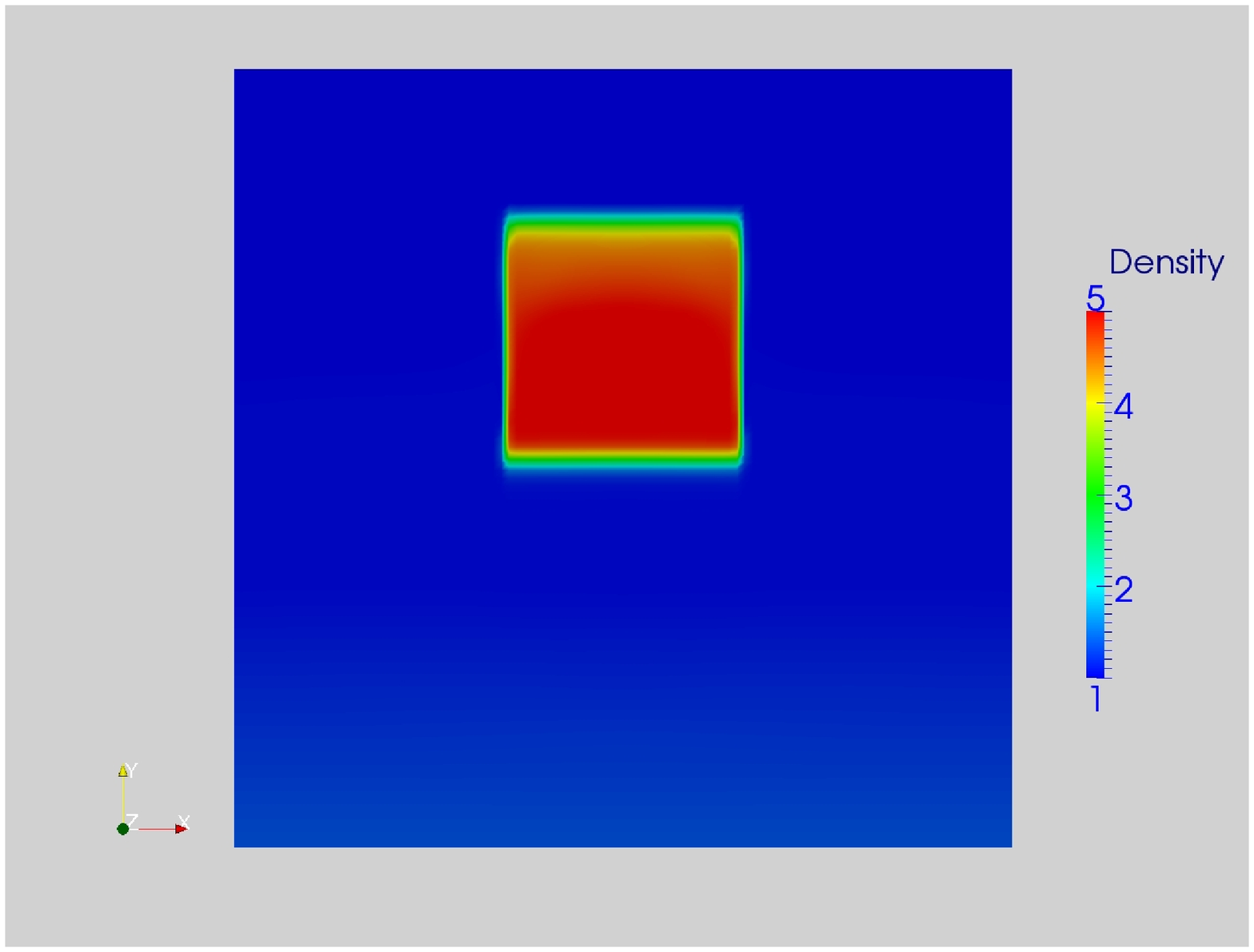}}
  \subfigure[$t = 0.2$ s]%
  {\includegraphics[width=0.49\textwidth]{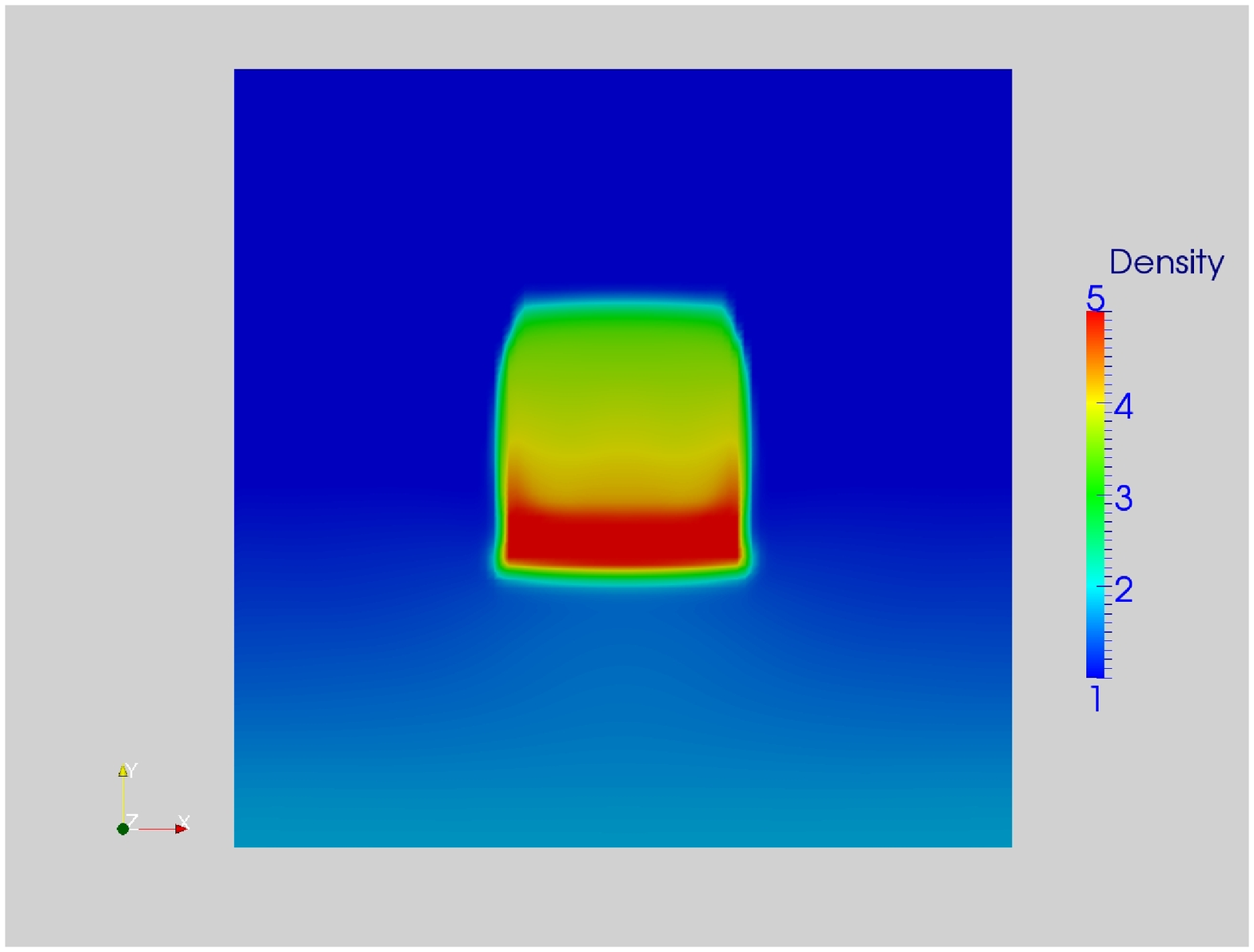}}
  \caption{Water drop test case: initial acceleration and deformation stage.}
  \label{fig:drop12}
\end{figure}

\begin{figure}
  \centering
  \subfigure[$t = 0.3$ s]%
  {\includegraphics[width=0.49\textwidth]{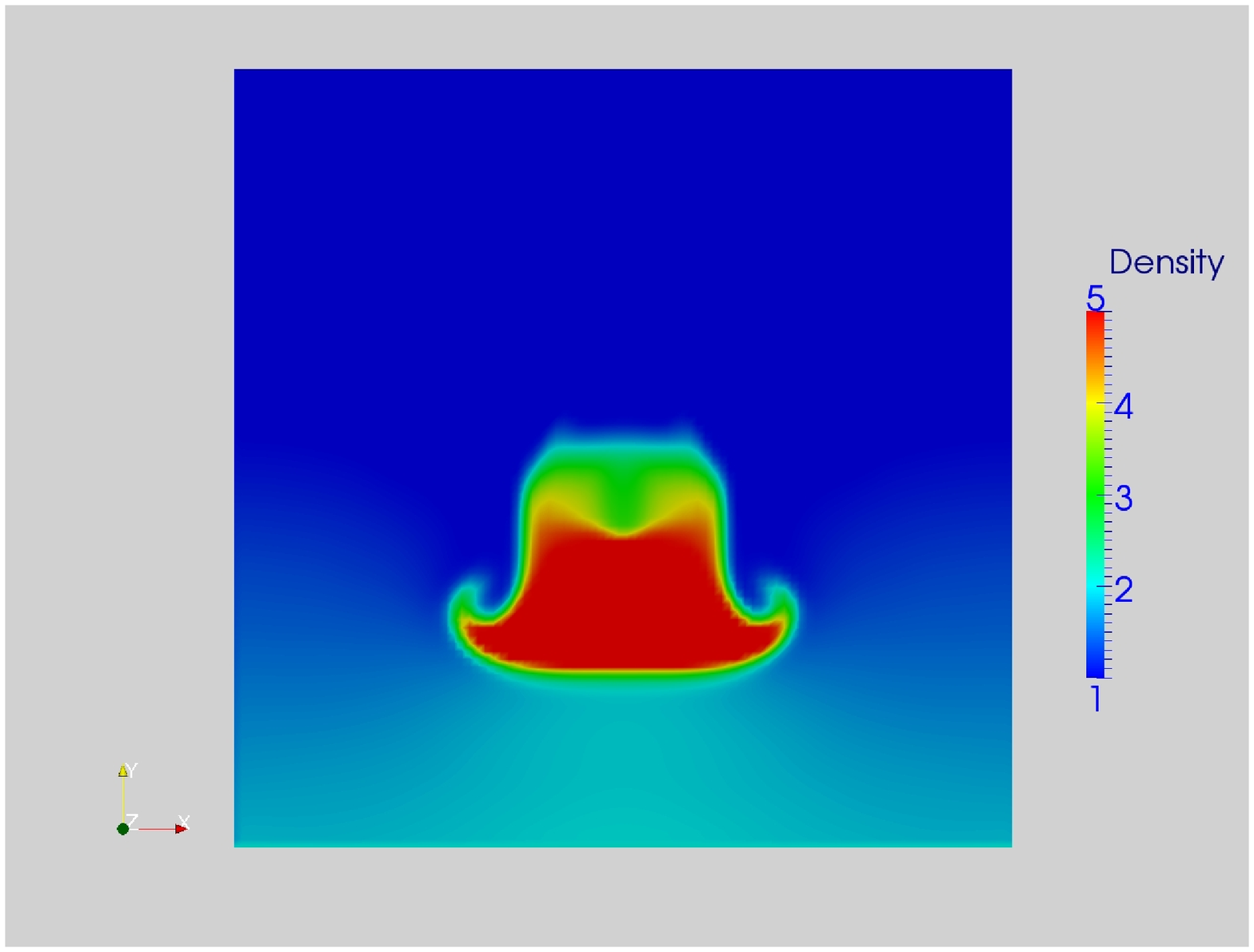}}
  \subfigure[$t = 0.4$ s]%
  {\includegraphics[width=0.49\textwidth]{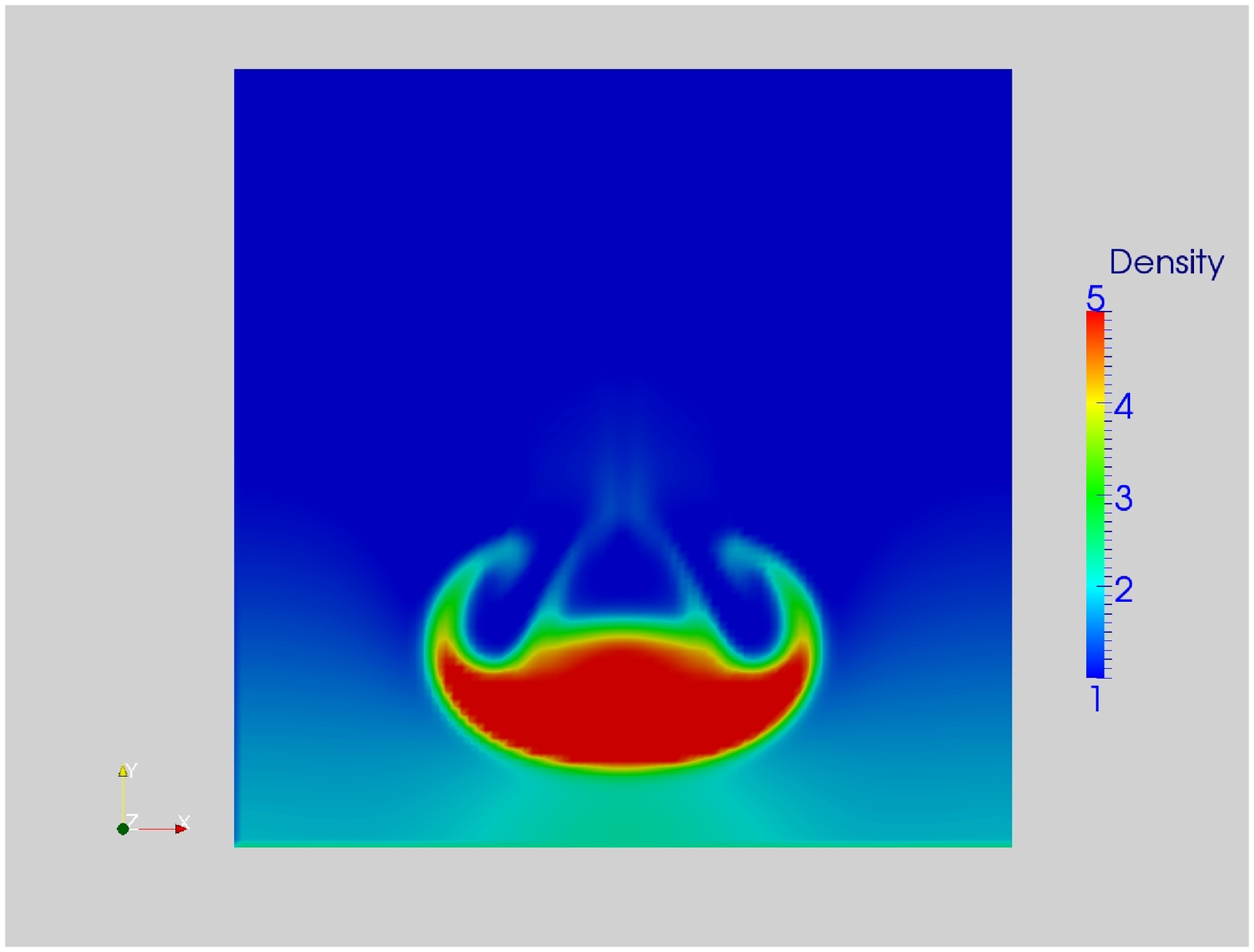}}
  \caption{Water drop test case: falling and further deformation of the drop.}
  \label{fig:drop34}
\end{figure}

\begin{figure}
  \centering
  \subfigure[$t = 0.5$ s]%
  {\includegraphics[width=0.49\textwidth]{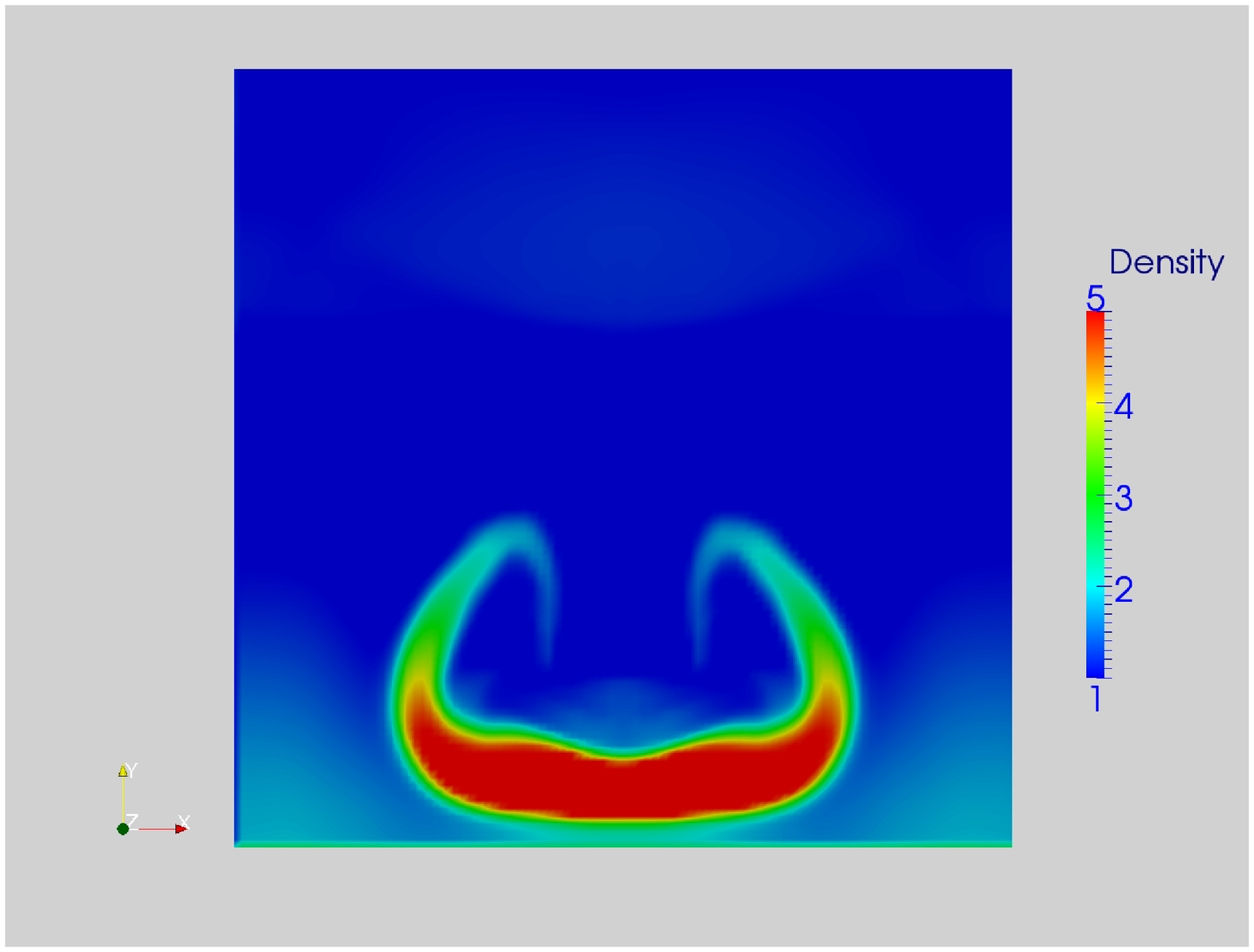}}
  \subfigure[$t = 0.6$ s]%
  {\includegraphics[width=0.49\textwidth]{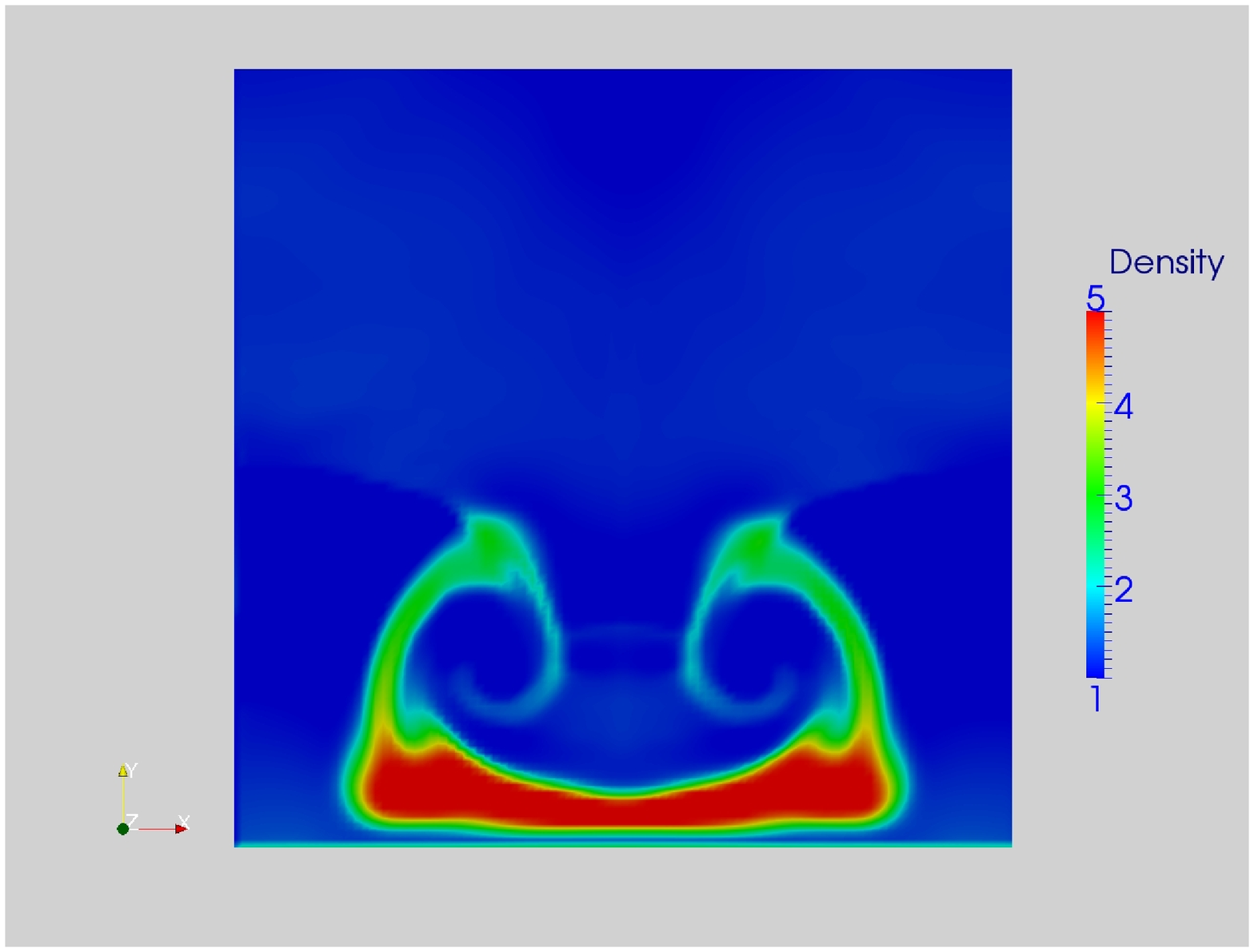}}
  \caption{Water drop hitting the tank bottom.}
  \label{fig:drop56}
\end{figure}

\begin{figure}
  \centering
  \subfigure[$t = 0.7$ s]%
  {\includegraphics[width=0.49\textwidth]{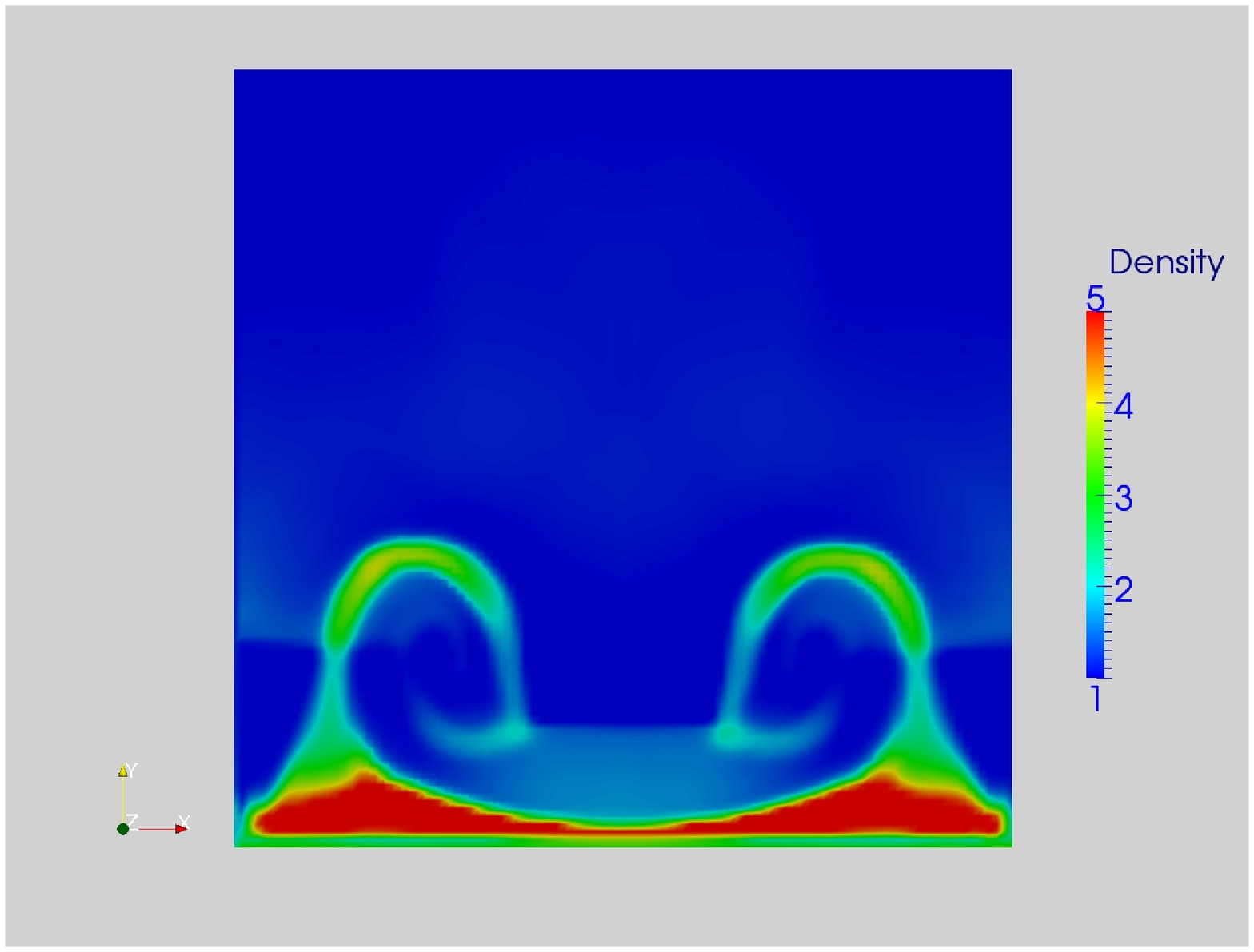}}
  \subfigure[$t = 0.8$ s]%
  {\includegraphics[width=0.49\textwidth]{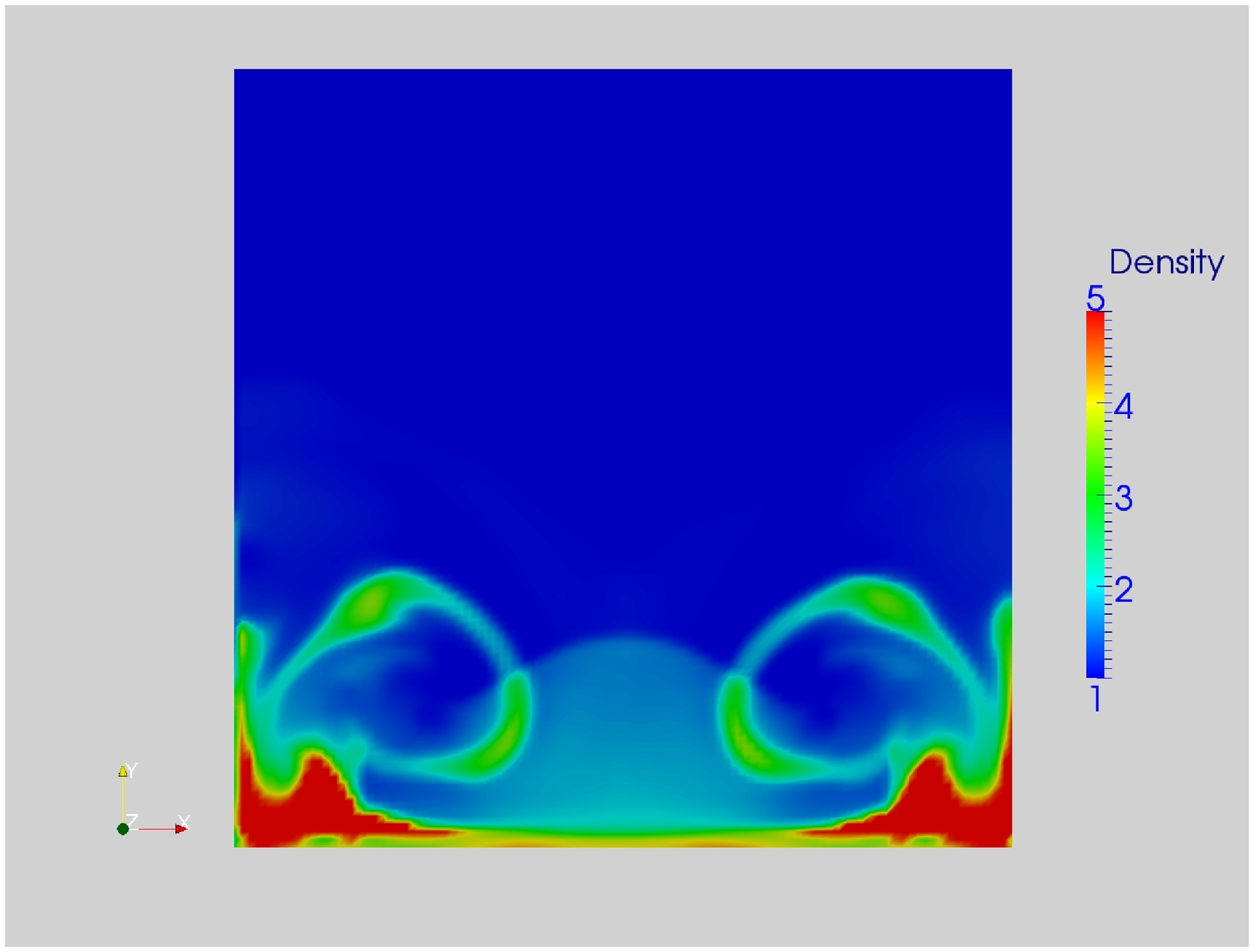}}
  \caption{Water drop flow on the bottom.}
  \label{fig:drop78}
\end{figure}

The sketch of this numerical experiment is given on Figure \ref{fig:drop}. The values of parameters are given in Table \ref{tab:eosDrop}. Initially the velocity field is taken to be zero and the pressure field is uniform in all domain $p_0 = 10$ Pa. The gravity force is taken to be $g = 10$ m/s$^2$ and the computational domain is discretized with a uniform grid of $100\times 100$ control volumes. Results of this simulation are presented in Figures \ref{fig:drop12} -- \ref{fig:drop78}. A similar test case has already been considered in \cite{Dutykh2007a,Dias2008,DDG2008} and results are in good qualitative agreement.

\begin{table}
  \begin{center}
    \begin{tabular}{r||c|c|c|c|c}
       & $\gamma^\pm$ & $\pi^\pm$, Pa & $c_v^\pm$, $\frac{J}{kg\cdot K}$ & $\rho^\pm$, $\frac{kg}{m^3}$ & $\alpha^+$ \\
    \hline\hline
    Heavy fluid, $+$ & 1.6 & 0.0 & 1.0 & 5.0 & 0.99 \\
     Light fluid, $-$ & 1.4 & 0.0 & 1.0 & 1.0 & 0.01 \\
    \end{tabular}
  \end{center}
  \caption{Equation of state parameters for the water drop test case.}
  \label{tab:eosDrop}
\end{table}

\section{Perspectives and conclusions}\label{sec:concl}

In this study we considered several two-fluid models. We began the exposition by the so-called six equations model (\ref{eq:mass6}) -- (\ref{eq:energy6}). Despite recent progress \cite{Aubry1995,Ghidaglia1996,Toumi1999,Ghidaglia2001,Bestion2005,Staedtke2005,Benkhaldoun2006,Rovarch2006,Ndjinga2008},\\ \cite{Galassi2009}, this system still represents some major difficulties for the numerical solution. Namely, the advection operator may be non-hyperbolic and contains non-conservative terms to be defined in some sense for discontinuous solutions. That is why, the six equations model was simplified through the velocity and energy relaxation process (see Section \ref{sec:relax}). In this way, we formally derived the so-called four equations model (\ref{eq:four1}) -- (\ref{eq:four3}) recently proposed as a model for violent aerated flows \cite{Dutykh2007a,Dias2008a,Dias2008,DDG2008}. Thus, the present work can be considered as an attempt towards further comprehension and at least formal justification of single velocity, single energy two-phase models. The resulting system (\ref{eq:four1}) -- (\ref{eq:four3}) is hyperbolic for any reasonable equation of state \cite{Dias2008} and possesses several nice properties. In particular, in Section \ref{sec:inv} we show that invariant regions $\alpha^\pm\in[0, 1]$ are preserved under the system dynamics. This property is necessary for the well-posedness of the system. In Section \ref{sec:Mach} we also formally derived the incompressible limit as the Mach number tends to zero. As a result, we recover usual two-fluid incompressible Navier-Stokes equations \cite{Popinet1999,Scardovelli1999,Chen1999} if both fluids are assumed to be Newtonian. Finally, several numerical results are presented in Section \ref{sec:num}.

\section*{Acknowledgement}

The support from the Research network VOR (Professors Jacky Mazars and Denis Jongmans) and Cluster Environnement through the program ``Risques gravitaires, s\'eismes'' is acknowledged.

We would like to thank our colleagues and friends Didier Bresch and C\'eline Acary-Robert for their continuous help and support. 

The second author thanks Professors Jean-Michel Ghidaglia and Fr\'ed\'eric Dias for introducing him to the beautiful field of two-phase flows.

\bibliographystyle{alpha}
\bibliography{biblio}
\end{document}